\documentclass[reqno,oneside, final, 12pt]{article}
\usepackage[russian,english]{babel}
\usepackage{latexsym,amsmath,amssymb,amsbsy,graphicx}
\usepackage[cp1251]{inputenc}
\usepackage[T2A]{fontenc}
\usepackage[space]{cite} % If exist.
\usepackage{hyperref}% add hypertext capabilities
\usepackage{color}
\usepackage{soul}

\makeatletter\def\@biblabel#1{\hfill#1.}\makeatother

\allowdisplaybreaks
\begin {document}
% ======================================

\noindent\begin{minipage}{\textwidth}
To be published in a special issue of the {\it Moscow University Physics Bulletin} dedicated to the 270th anniversary of the University's founding. 
\begin{center}

% {\large ФИЗИКА ... \small(рубрика прописными)}\\[1cc]
% Рубрика обычно указывается редактором.
{\Large{Field structures and singularities in subwavelength optics}}\\[9pt]

{\large M.I. Tribelsky$^{1a}$, B.S. Luk'yanchuk$^{2b}$}\\[6pt]

% Цифровые индексы указываются, если у авторов разные почтовые адреса;
% буквенные -- если авторов несколько, для всех авторов, отвечающих на корреспонденцию.
% Ниже и в сведениях на английском -- аналогично.

\parbox{.96\textwidth}{\centering\small\it
$^1$ Chair of Polymer and Crystal Physics, Faculty of Physics, M.V. Lomonosov Moscow State University, Russia, 119991, Moscow, Leninskie Gory 1, Bldg. 2.\\

$^2$ Chair of Nanophotonics, Faculty of Physics, M.V. Lomonosov Moscow State University, Russia, 119991, Moscow, Leninskie Gory 1, Bldg. 35.\\
\ E-mail: $^a$mitribel@gmail.com,
$^b$LukyanchukBS@my.msu.ru}\\[1cc] % не более 2 e-mail адресов в шапке статьи!!!!

\parbox{.96\textwidth}{\centering\small Received --.--.2024, Accepted --.--.2024.} % Даты указываются редактором.
\end{center}

{\parindent5mm 
A brief overview of the current state of the problem of electromagnetic field singularities arising from the refraction and scattering of light by material objects is given. The discussion begins with caustics arising from ray tracing in geometric optics and consistently moves toward increasing the accuracy of consideration and decreasing the scale, ending with a description of singularities in light scattering by subwavelength particles. Common and distinctive features of various types of singularities, the role of the symmetry of the problem and the law of conservation of energy are revealed. Physical foundations and methods for overcoming the diffraction limit are discussed. The theoretical description is illustrated by experimental examples. Various practical applications of the effects under consideration are indicated.}
\mbox{}\vspace{1pt}\par

\small UDK: 535.31; 535.36; 535.4 \vspace{1pt}\par
\vspace{1pt}\par
\end{minipage}
% ======================================

% Команды автора:
% \newcommand...
% (Просьба указывать только используемые в настоящей статье.)

\definecolor{vs}{rgb}{0.1,0.4,0.1}                  % dark green
       % \del{...} to delete the text "..."

\section*{Introduction}% \small\rm(не нумеруется)}
\mbox{}\vspace{-\baselineskip}
\subsection*{Preface}

Academician V.E. Zakharov noted that Russian science is a unique cultural phenomenon that emerged thanks to the reforms of Peter the Great. By the time of Peter's death in 1725, Russia already had an Academy of Sciences, but science in the modern sense of the word was practically non-existent. By this time, Western Europe had (if we count from Copernicus) almost two hundred years of continuous development of essentially modern science. Russian science was able to overcome this time lag, and at the beginning of the 20th century it was already quite at the level of European science. The best evidence of this is that one of the first Nobel Prizes in 1904 was awarded to a Russian scientist --- Ivan Petrovich Pavlov. This was a prize in the field of medicine, but in physics at the end of the 19th -- beginning of the 20th century, Russian scientists also made world-class discoveries.

A statue of A.G. Stoletov, the first Russian physicist to receive European recognition, is installed in front of the entrance to the Faculty of Physics of Moscow State University. Between 1862 and 1866, Stoletov completed an internship abroad, where Gustaf Kirchhoff called him his most talented student. Stoletov organized the first educational and research physics laboratory in Russia. He also initiated the creation of the Physics Institute at the Imperial Moscow University {(this was the official name of MSU until 1917). Most of the physics professors at IMU} were his students. After the death of A.G. Stoletov, the Physics Department was headed by Professor Nikolai Alekseevich Umov since 1896, who in 1874 first introduced the concept of the speed and direction of movement of the energy flow. In 1884, J. Poynting applied the concept of energy flow density introduced by Umov to the electromagnetic field, defining a vector that is sometimes called the Umov-Poynting vector. In front of the building of the Faculty of Physics of Moscow State University there is also a monument to Pyotr Nikolaevich Lebedev --- an outstanding experimental physicist who was the first to measure the pressure of light.

Wars, revolutions and the events that followed them have caused enormous damage to the development of science in Russia, but have not stopped it. In general, the ``golden age'' of Soviet science, when in a historically short period of time, in a few decades, it covered the distance that took Europe several centuries, appears to be an effect caused by positive feedback --- a phenomenon like a thermal explosion. It is practically impossible to repeat this experience, as well as to reconstruct the unique educational system in the Soviet Union, which, in the second half of the 20th century, was possibly the best in the world (at least in the field of mathematics and natural sciences).

The authors of this review know this first-hand. The years of our formation as theoretical physicists coincided with the period when Moscow was rightly considered one of the world capitals of theoretical physics, and foreign scientists studied Russian in order to be able to read the articles of their Soviet colleagues in the original. Unlike the ``elderly Negro'' who ``would have learned Russian only because Lenin spoke it\footnote{A verse by famous Russian-Soviet poet Vladimir Mayakovsky.}'', the motivation of foreigners was connected with the fact that they received {Russian-language versions of Soviet scientific journals at their universities significantly faster than their English-language translations.} Moreover, not everything was translated, and the quality of the translations left much to be desired.

Moscow University played a vital role in forming many sections of Russian science, {and one of the most significant is the contribution of MSU scientists to the creation of} the Russian optical school, {especially} the section ``nonlinear optics''. Even this term was introduced into science by our compatriots S.I.~Vavilov and V.L. Levshin. The names of Russian scientists L.I. Mandelstam, A.A. Andronov, G.S. Landsberg, S.M. Rytov, M.A. Leontovich; {in more recent times, R.V.~Khokhlov, L.V. Keldysh, A.S. Akhmanov, A.P. Sukhorukov, A.S. Chirkin}, and many, many others speak for themselves and characterize the significance of this contribution. For this reason, the authors, who are both alumnae of the Physics Faculty of Moscow State University, are especially pleased to publish this review in the special issue of the Moscow University Bulletin, dedicated to the \mbox{270th anniversary} of our Alma Mater, as a tribute to our native university and our teachers.

\subsection*{Introduction itself} 

{This review is intended for a wide range of readers and is devoted to discussing field structures and singularities in optics. This subject belongs to a large and rapidly developing area. Since ``it is impossible to embrace the immensity,'' the review does not pretend to be complete in any way. We have selected only a few simple cases for discussion and concentrated on clarifying the physics of the phenomena under consideration. At the same time, many important issues, both from a fundamental point of view and the point of view of various applications occur outside the scope of the review. It is partly compensated by many references that address the relevant publications to the interested reader.}

When discussing singularities, we will move towards decreasing characteristic scales --- from geometric optics to wave optics and, then, subwavelength optics. Here, it is necessary to define what we mean by the scale of a singularity since usually the corresponding structure's characteristic scale vanishes when approaching a singularity. Therefore, speaking about the scale of a singularity, we mean its {\it maximal scale,} i.e., the characteristic size of the region within which the field structure is mainly determined by the type of the singularity.

In {theoretical analysis of the electromagnetic field}, the Gaussian system of units is used, in which the electric (E) and magnetic (H) fields have the same dimension. The time dependence of a monochromatic wave is assumed to be of the form $\exp \left( { - i\omega t} \right).$ In addition, all substances will be considered non-magnetic, since the magnetic permeability $\mu$ at optical frequencies equals unity~\cite{LL_Electrodyn}, and the medium in which the light-scattering objects are located is non-dissipative. Unless otherwise specified, the latter is a vacuum.

The review is structured in the following manner. It begins with a brief history of the issue in Section~\ref{sec:History} Section~\ref {sec:GO} discusses caustics, followed by Section~\ref {sec:WO} devoted to diffraction catastrophes and related phenomena. Section~\ref{sec:phase_sing} moves on to a discussion of phase singularities, optical vortices, superoscillations, and photonic jets. Section~\ref{sec:Poynting} is devoted to singularities of the Poynting vector field. The {Conclusion} briefly summarizes the results, providing a clear overview of the review's content.

\section{History of the issue\label{sec:History}}

The history of optics is connected with studying light structures of various scales {and also moved towards decreasing the latter.} Astronomical and geophysical optical phenomena (sunrises, sunsets, rainbows, etc.) were known even to primitive man. 
Several light structures of smaller scales have long been observed in glass products (glass already existed in Egypt 5 thousand years ago). Nero's tutor Seneca, reported on the magnifying property of glass spheres~\cite{Seneca}. By the way, Emperor Nero himself watched gladiator fights through a specially cut emerald\footnote{There is a hypothesis that Nero suffered from a visual defect, and the emerald was just glasses, but certainly one of the first.}. In his {\it Natural History\/}, Pliny the Elder described the incendiary effect of glass spheres~\cite{Pliny}. The description of these phenomena in optics is carried out with the help of geometric optics, operating with the representation of a family of rays introduced into science by Euclid. Although even ancient scientists (Archimedes, Ptolemy, etc.) knew the effect of light refraction, it took a whole millennium to establish the exact law, known now as Snell's law. The law was discovered in Baghdad by the Persian scientist Abu Said al-Ala Ibn Salem in 984~\cite{kwan2002a}. Subsequently, this law from 1602 to 1637~ was rediscovered by Thomas Harriot, Johannes Kepler, Willebrord Snellius, and Ren\'e Descartes. In 1662, Pierre de Fermat {established} that this law follows from Fermat's principle, according to which light follows the path that minimizes its transit time. In 1678, Christiaan Huygens showed how Snell's law can be explained using the wave nature of light and the Huygens-Fresnel principle.

Theoretical description of light structures is carried out at different levels: light can be described as rays, scalar waves, vector fields, and quantum fields~\cite{berry2023a,Bliokh2023,Shi2024}. Using the terms scalar and vector fields in optics requires clarification --- after all, the electromagnetic field always consists of vector fields $\vec{E}$ and $\vec{H}$. The point is, however, that in some cases, all components of these fields are expressed through a single scalar function satisfying the wave equation. These electromagnetic fields are called scalar. The simplest example of such a kind is a monochromatic plane wave propagating in free space. The corresponding electromagnetic field is usually called a vector field if such a function does not exist.

Berry's review~\cite{berry2023a} considers singularities arising from different descriptions of the electromagnetic field: if the propagation of electromagnetic radiation is described by rays, as is done in geometric optics, then the singularities are caustics; in scalar wave optics, these are phase singularities; in vector waves, these can be singularities in which the polarization of light is purely linear or purely circular, {etc.} The emphasis in review~\cite{berry2023a} is on the typicality, structural stability, and universality of the corresponding singularities.

In this review, we apply these general principles to the description of subwavelength-scale light structures. In modern optical devices, such structures are realized through various methods for overcoming the diffraction limit. The diffraction limit, associated with the name of Ernst Abbe~\cite{Abbe1873,Hon.1882}, actually, arises due to the Heisenberg uncertainty principle for the momentum and coordinates of a photon: {$\Delta x\,\Delta {p_x} \leq \hbar/2$}. According to quantum mechanics, for a photon propagating in a medium with a refractive index $n$, the relation $p=n\hbar k$ is valid, where $k=\omega/c$ and $c$ are, respectively, the wave number and the speed of light in a vacuum. This allows us to obtain a lower bound on the accuracy of measuring the coordinates of a photon: {$\Delta x\, \leq \lambda /(2n)$~\cite{novotny2012a}.} However, it's crucial to note that this is a lower bound, as the quality of the observation system can significantly influence the actual resolution. If the quality is not high enough, the resolution will be determined by the properties of the optical device, not the fundamental estimates given by the diffraction limit.

For a significant period, the diffraction limit was perceived as an insurmountable barrier. However, approximately a century ago, a revolutionary concept of near-field microscopy emerged. The concept employs the unique characteristics of light passing through a small aperture. This method's resolution could surpass the diffraction limit by a significant margin. The brilliance of this idea lies in its simplicity ---  when the aperture size is much smaller than the wavelength, this scale (not the wavelength) acts as a characteristic scale for the light beam's intensity variations in the near field.

However, what about the uncertainty principle? There is nothing wrong with it. For a photon passing through a small aperture, its interaction with the diaphragm forming the aperture must be explicitly considered. Then, the relation $p=n\hbar k$ is not valid, and the above estimate of the diffraction limit (valid for a photon propagating in free space) is not applicable.

The introduction of near-field microscopy (SNOM) marked a pivotal moment in the field of optics. Subsequently, a range of other physical concepts were proposed to tackle the diffraction limit, reflecting the diverse approaches in the field. The current status of this issue is encapsulated in the comprehensive work~\cite{hecht2000a}. As for subwavelength optical microscopy, it has evolved into an independent field known as 'nanoscopy'~\cite {wang2011a,wang2016a,krivitsky2013locomotion,huang2018planar,astratov2023a,Maslov2019nanoscopy}.

It is interesting to note that the presence of the diffraction limit has another explanation, and, as often happens in physics (and not only in physics), the emergence of an alternative explanation for a problem leads to an alternative way to solve it. This other explanation is based on the fact that information about small (relative to the wavelength) spatial inhomogeneities of the field is transmitted by rapidly decaying evanescent waves. Their attenuation can be compensated for by super-oscillations in special antennas, in which part of the radiated energy is directed into small solid angles~\cite{berry2019roadmap}. Surprisingly, the theoretical foundations of the theory of super-oscillations were laid by the work~\cite{oseen1922einsteinsche}, which introduced the concept and mathematical framework of super-oscillations, more than a hundred years ago, see also~\cite{di1952super} and others. Nonetheless, experimentally, super-oscillations of electromagnetic fields have only begun to be intensively studied in recent years; see \cite{berry2019roadmap} and the works cited therein. We will return to the discussion of the nature and properties of superoscillations in more detail below; see Section \ref{sec:phase_sing}

{In conclusion, in this section, we will dwell on the classification of optical phenomena depending on the ratio between the radiation wavelength $\lambda$ and the characteristic scale of the problem $R$. Quantitatively, this ratio is usually characterized by the value of the dimensionless quantity $q = kR$, called the {\it size parameter}. In classical optical devices, this parameter is huge. For example, for visible light $k \sim 10^5$ cm$^{-1}$, which for an aperture size of about 1 cm gives $q\sim 10^5$. It is the field of {\it geometrical optics}. Another limiting case $q \ll 1$ corresponds to {\it Rayleigh} and also {\it anomalous\/} scattering (for a discussion of the latter, see, for example,~\cite{tribelsky2006a,TribelMirosh_UFN_2022}). The intermediate case $q\sim 1$, when the first few Mie resonances (dipole, quadrupole, octupole) play the dominant role in the scattering of light by material objects, is sometimes called the {\it Mie-tronics} region~\cite{kivshar2022Mie-tronics}. Recently, the attention of researchers has also been attracted to the study of light scattering by objects with $q\sim 10$, for which the term {\it mesotronics}~\cite{minin2021mesoscopics} is beginning to be used.}

\section{Geometric optics, caustics \label{sec:GO}}
\mbox{}\vspace{-\baselineskip}

Let us begin the discussion of singularities with geometric optics (GO). Note that the term GO is not entirely unambiguous. Initially, GO arose as a description of optical phenomena using ray tracing. Later, it was realized that GO corresponds to a certain approximation in solving Maxwell's equations, where the phase of the wave plays an important role. Following the issue's history, we will limit ourselves to ray tracing in this section.

What happens when a parallel beam of rays falls on a homogeneous transparent sphere with a refractive index $n>1$? Using ray tracing~\cite{Arnold1990} and Snell's law (a method known since Kepler), we can verify that the rays refracted at the boundary of the sphere form a caustic created by a family of intersecting rays being their envelope, see Fig.~\ref{fig:cusp}.

The caustic in Fig. \ref{fig:cusp}(b) is described by a system of parametric equations~\cite{luk2022a}:  \begin{equation}\label{eq:cusp} {x_{\text{c}}} = \left[ {1 - \frac{1}{2}\frac{{\sqrt {{n^2} - 1 + {{\cos }^2}\varphi } - 2\cos \varphi }}{{\sqrt {{n^2} - 1 + {{\cos }^2}\varphi } - \cos \varphi }}\cos \varphi } \right]\cos \psi;\;\;\; {y_{\text{c}}} = \sec \psi \,\sin \varphi - {x_{\text{c}}}\tan \psi,
\end{equation}
where $\psi = 2\left[ \varphi - \arcsin \left( \frac{\sin\varphi}{n}\right)\right]$. Caustics obtained by irradiation with a point source look similar~\cite{zalowich2001a}.

The type of caustic shown in Fig.~\ref{fig:cusp} in the two-dimensional case (cylindrical lenses) is called a cusp, from the English word cusp, which according the Merriam-Webster Dictionary definition means {\it a fixed point on a mathematical curve at which a point tracing the curve would exactly reverse its direction of motion}. In the three-dimensional case, the caustic is a surface. For the situation shown in Fig.~\ref{fig:cusp}, this surface is obtained by rotating the cusp around the axis of symmetry passing through the center of the sphere. Such a surface is called a cuspoid. In the neighborhood of the cusp's (cuspoid's) tip, its shape is universal for all caustics of this type and is described by a semi-cubical parabola:

\begin{equation}\label{eq:Semiqube}
y^2+(a x)^3=0;\;\; a=const
\end{equation}
where $x$ and $y$ are measured from the cusp's tip. From formula (\ref{eq:cusp}), this expression can be obtained by expanding the $x_c(\varphi)$ and $y_c(\varphi)$ dependencies in small $\varphi$, and excluding $\varphi$ from the resulting relations.

Assuming that the intensity inside the caustic cone is homogenized, {for example, due to local imperfections in the optical properties of the radiation-focusing sphere and its surface, one can find the intensity gain coefficient at the caustic output from the sphere's \mbox{surface~\cite{arnold2003b,Luk2003_Cleaning}:}
\begin{equation}\label{eq:cusp_intens}
\frac{S_{\max }}{S_0} \approx \frac{27{n^4}}{\left( 4 - {n^2} \right)^3}.
\end{equation}
Here $S_0$ is the intensity (energy flux density) of a plane wave incident on a sphere, and $S_{\max}$ is the intensity averaged over the area of a circle bounded by the line of intersection of the caustic cone with the surface of the sphere.}

The formation of caustics explains the igniting action of glass balls, described by Pliny the Elder~\cite{Pliny}. In addition to the cusp, there are other types of caustics. They are classified using the theory of catastrophes~\cite{Arnold1990,Arnold_UFN_1983,Arnold1996,Arnold_UFN_1999,berry2023a,berry1980iv,kravtsov1993a}.
Caustics occur in many optical (and not only optical) phenomena, such as rainbows, mirages, distortion of images in curved mirrors, etc.~\cite{berry2023a,Arnold1990,berry1980iv,kravtsov1993a}. A rapidly changing light ``network'' can be observed on a sunny day at the bottom of a shallow basin with small waves on its surface is also a set of caustics.

In conclusion of this section, we note that although GO refers to the coarsest level of description of light structures in optics~\cite{berry2023a}, it works well in a wide range of parameters and has a transparent physical meaning, explicitly showing the direction of light energy flows.

\section{Wave optics, diffraction catastrophes, the Pearcey integral and its generalizations, photonic jets \label{sec:WO}}
\mbox{}\vspace{-\baselineskip}

Though the Geometric Optics (GO) approach is effective for describing optical phenomena, it is only an approximate solution of Maxwell's equations. It is valid if the characteristic scales of the corresponding electromagnetic field structure are much larger than its wavelength. The actual applicability threshold depending on the refractive index ($n$) and the shape of the scattering particle. For simple shapes like spheres or cylinders at $1 < n < 2$, the size parameter $q=kR$, where $R$ is the sphere (or cylinder) radius, should be several tens for GO to provide reasonable accuracy. However, regardless of the size parameter value, GO becomes inapplicable near singularities, as the characteristic spatial scale of the field variations becomes comparable to (or even smaller than) the wavelength of the radiation. Then, dealing with caustics requires a higher accuracy than that GO can provide. It can be achieved with the help of wave optics. Specifically, this is true for caustics. In analogy with GO, where caustics are described by catastrophe theory, in wave optics, they are referred to as ``diffraction catastrophes,'' which correspond to sharp coordinate dependences of the intensity distribution of a wave field~\cite {trinkaus1977,berry1980iv,berry1996colored, kofler2006a}.

Let us illustrate this with an example. Consider a wavefront converging to a caustic. We choose an arbitrary point on the front surface and draw the two mutually orthogonal planes of the principal normal sections passing through this point (the planes, where the principal radii of curvature lie). Let us construct around the chosen point an infinitesimal element of the wavefront surface area in the form of a curvilinear  rectangle whose sides are parallel to the lines of intersections of the specified planes with the wavefront surface; see Fig.~\ref{fig:Contraction}. Let us draw the rays emanating from each point of the perimeter of this rectangle (such a construction is called a ray tube). In GO, the direction of the energy flow coincides with the rays, which are always orthogonal to the wavefront surface. Therefore, the energy flux through the sidewalls of the ray tube is zero, and the energy flux through any of its sections is constant. However, as the wavefront approaches the caustic, the area of the corresponding curvilinear rectangle decreases; see Fig. ~\ref{fig:Contraction}. Considering that the total flux must be conserved, the energy flux density increases as the wave front approaches the caustic. At the same time, the radii of curvature of the wavefront decrease, i.e., its curvature increases. Eventually, the rays lying in the plane corresponding to the smaller value of the radius of curvature ($R_1$ in Fig.~\ref{fig:Contraction}) intersect, i.e., form an element of the caustic, and the rectangle degenerates into a line segment. This line in Fig.~\ref{fig:Contraction} is designated $C_1$. In the GO approximation, the radiation intensity on this line becomes infinite, meaning that the energy flux density at this point is exceptionally high. It is a key implication of the wavefront approaching a caustic. In the three-dimensional case under discussion, the set of these lines forms the surface of the caustic (in the two-dimensional case, this is a line).

However, this is not the end of the story. Having passed through $C_1$, the wave in the plane coinciding with the plane of the figure transforms from converging to diverging. In contrast, in the plane perpendicular to the figure, the wavefront continues to converge and collapses into line $C_2$. The set of such lines constitutes another branch of the caustic. In the degenerate case, when $R_1=R_2$, both radii of curvature vanish at the caustic. It corresponds to the ideal focus of a locally spherical converging wavefront.

Thus, in GO, caustics are singular surfaces (lines) on which the radiation intensity diverges, and at least one of the principal radii of curvature of the wavefront vanishes (in the two-dimensional case, such a radius is unique). Let us also emphasize that, as noted above, the passage of a beam through a caustic is accompanied by a change in the sign of the radius of curvature of the wavefront --- a converging wave turns into a diverging one, which leads to the corresponding changes in the phase of the light wave~\cite{Kravtsov1980,kravtsov1993a}. However, wave optics shows that, although a significant increase in intensity occurs in the caustic region, it remains finite. In this case, the caustic ``blurs'' into a region of finite thickness.

As an example, Fig.~\ref{fig:Cyl_Cusp} shows the intensity distribution when a transparent cylinder focuses a plane electromagnetic wave. It is clearly seen that the caustic is blurred into a high-intensity region that has both a finite width and a specific internal structure associated with the interference modulation of the radiation intensity. A typical pattern of alternating light (constructive interference) and dark (destructive interference) regions is observed in its vicinity.

For a sphere and an infinite circular cylinder, both homogeneous and having a layered structure of the same symmetry, the diffraction problem admits an exact solution~\cite{bohren2004a}. Such solutions are usually called Mie solutions. We will also adhere to this terminology, although, strictly speaking, the Mie solution itself refers to the scattering of a plane linearly polarized wave by a homogeneous sphere~\cite{Mie1908}. In these solutions, the scattered radiation is represented as an infinite sum of partial waves: dipole, quadrupole, octupole, etc. (the so-called {\it multipole expansion}). Such a series quickly converges for a small radius {of the light-scattering} particle. The larger the radius, the slower the convergence, and the greater the number of partial waves that must be taken into account to ensure a given calculation accuracy. Therefore, approximate calculation methods have been developed for particles whose size is large compared to the radiation wavelength. In addition to being more convenient in practice, they can be used to clarify the physics of the phenomenon hidden in the Mie solution due to the distribution of the scattered field over a large number of partial waves. 

Thus, for example, in the simplest case of focusing a plane, linearly polarized wave, incident normally on a cylindrical lens whose axis is parallel to the plane of polarization, the distribution of fields in the vicinity of the caustic is approximately described by the so-called Pearcey integral~\cite{Pearcey1946}.
\begin{equation}\label{eq:Pearcey}
 I_P(X,Z)\;\; = \;\;\frac{1}{\sqrt {2\pi}}\int_{- \infty}^{\infty}e^{-i\left( {X\,{\xi}\; + \;Z\frac{{\xi^2}}{2}\; + \;\frac{{\xi^4}}{4}} \right)}\,d{\xi}.
\end{equation}

Here, the coordinate origin coincides with the caustic tip in the GO approximation, the wave propagates along the $z$-axis, and the axis of the cylindrical lens is located along the $y$-axis. In addition, in (\ref{eq:Pearcey}) new dimensionless variables \mbox{$X=const\cdot xk^{3/4}$} and \mbox{$Z=const\cdot z\sqrt{k}$ were introduced.}

Supplementary to the optical problems we discussed, this integral plays an essential role in other wave phenomena (hydrodynamic and acoustic waves, quantum mechanics, etc.~\cite{Connor1981}). Therefore, it has been studied in sufficient detail. In particular, although it is not taken in quadratures, many of its analytical approximations are valid in various ranges of values of the variables $X$ and $Z$; see, for example, the work~\cite{Lopez2016}.

Generalizing this approach to the case of axially symmetric beams, instead of the Pearcey integral, we arrive at integrals of the Bessoid type:
\begin{equation}\label{eq:Bessoid}
  I_m(\rho,Z)=\int_{0}^{\infty}\xi^{m+1} J_m(\rho\xi)e^{-i\left( {Z\frac{{\xi^2}}{2}\; + \;\frac{{\xi^4}}{4}} \right)}\, d{\xi},
\end{equation}
where $m$ is a non-negative integer, $J_m(\zeta)$ is the corresponding Bessel function, and \linebreak \mbox{$\rho = const\cdot rk^{3/4}$.} Here, $r$ is the distance from the beam axis to the observation point.
Using these integrals makes it possible to describe not only paraxial\footnote{Recall that paraxial beams are the ones characterized by small angles of deviation of the rays from both their axis and the normals to the refracting surfaces.} scalar beams, but also strongly non-paraxial axially symmetric beams, including vector beams~\cite{kofler2006a}. An example of using such an approach is shown in Fig.~\ref{fig:Nikita}, which shows the intensity of the field arising when a plane linearly polarized wave passes through a transparent sphere~\cite{kofler2006a}.

From a practical standpoint, the most significant outcome of this analysis is the determination of the shift of the maximum field intensity (the diffraction focus) in relation to the position of the geometric focus predicted by GO and located at the tip of the caustic. The authors~\cite{kofler2006a} derived a straightforward approximate formula that describes the position of the diffraction focus:
\begin{equation}\label{eq:Fdiff}
  f_d \approx \frac{R}{2} \frac{n}{n-1}\left(1-\sqrt{\frac{3 \pi}{4 q} \frac{n(3-n)-1}{n(n-1)}}\right) .
\end{equation}
Here, $f_d$ is the distance from the sphere center to the diffraction focus point, and the factor before the brackets is the same, but for the focus position predicted by GO. Thus, the second term in the brackets describes the specified shift. Note, the shift is always negative --- the diffraction (i.e., {\it actual}) focus is always situated nearer to the surface of the sphere than the position of the caustic tip predicted by the GO approximation.

When comparing this description of wave phenomena with the results of the exact Mie solution, it becomes evident that for $q \sim 100$ (which in the optical range corresponds to a micron-size sphere) and $1.4<n<1.6$, the formula (\ref{eq:Fdiff}) provides an accuracy of no worse than 5\% \cite {kofler2006a}. As for the approach to the problem itself, in the range of $n$ values corresponding to different optical glasses, this approximation works well for $q>100$ and more or less satisfactorily in the range of $30\leq q \leq 100$. The angular orientation of the plane of the cross-section of the resulting field has little effect on the intensity distribution in this plane (compare the upper and lower panels in Fig. \ref{fig:Nikita}). The departures from the Mie solution increase for smaller size parameter values. However, in this region, the multipole expansion used in the Mie theory becomes rapidly convergent, and there is no need for an alternative description of the scattering process. Changing the size parameter towards its increase takes us at $q\gg 100$ into the area of technical physics (lens optics, etc.), which lies outside the scope of this review.

Returning to the discussion of the effects of visible light interaction with subwavelength objects, we note the complex structures that arise in the intensity distribution when a plane wave is scattered by a spherical particle lying on a flat substrate, see Fig.~\ref{fig:OnSubstrate_ Normal}. This problem, which is essential for many applications (for example, in laser cleaning of surfaces~\cite{arnold2003b,Luk2003_Cleaning,luk2000a,luk2002a,luk2004a,kane2006a,arnold2008_Cleaning}), also admits an exact solution~\cite{bobbert1986a}.

Interesting effects are observed when the incident wave deviates from the normal to the substrate surface. In this case, the problem loses its mirror symmetry, and in the GO approximation, a shadow region appears. Consideration of such a problem within the framework of wave optics shows that bright spots corresponding to a high concentration of electromagnetic energy can form in this region~\cite{luk2004a}; see Fig.~\ref{fig:Sphere_substrate_45}(a). These spots have been observed experimentally; see Fig.~\ref{fig:Sphere_substrate_45}(b).

Returning to the discussion of wave effects in the caustic zone, note a narrow, extended region of high radiation intensity near the caustic tip. Such a region is distinctly  visible in Fig.~\ref{fig:Cyl_Cusp}. It is called the {\it photonic jet.} The photonic jet arises in many problems of wave and subwavelength optics~\cite{Heifetz2009,Wang2022,Minin2015}. For example, it is clearly expressed in Fig.~\ref{fig:OnSubstrate_ Normal}, where it comes out of the ``crown'' of the irradiated sphere. Note that in this case, due to interference with radiation reflected from the substrate, the photonic jet is oriented opposite to the direction of propagation of the incident radiation.

Under certain conditions, the width of this jet becomes smaller than the diffraction limit~\cite{zhang2021a,littlefield2021a,minin2020a,minin2022a,chen2004a}. In this case, the physical basis for overcoming the diffraction limit is the same as that discussed above --- the features of near-field optics in the immediate vicinity of a material object.

{The ability to control the characteristics of the photonic jet has made it widely used in various fields, such as the previously mentioned laser cleaning {and processing} of surfaces~\cite{arnold2003b,Luk2003_Cleaning,luk2000a,luk2002a,luk2004a,kane2006a}; micro- and nanoelectronics~\cite{wu2007a,mcleod2008a,kim2012a,zhang2019a}; nano-detection of material objects~\cite{chen2020subwavelength}; medicine~\cite{astratov2010a,yan2020a}, etc. As an example of such an application, we present the technology of forming nanoholes in a surface covered with spherical microparticles~\cite{lu2000a,munzer2002a,huang2002a}. Fig.~\ref{fig:Holes} shows a view of such a surface with an array of nanoholes formed by the action of a laser pulse on a substrate covered with a lattice of $SiO_2$ micron-size spheres. Despite the micron size of the spheres, due to the formation of photonic jets, {as well as the essentially nonlinear absorption and ablation processes initiated by them}, it is possible to form nanometer-size holes in the substrate (up to 30 nanometers in diameter~\cite{huang2002a,huang2002b}).}

Nanometer-scale structures exhibiting notable subwavelength sizes can be generated on a substrate's surface through the laser irradiation of plasmonic nanoparticles --- metal nanoparticles characterized by a negative real part of permittivity (Re$\,\varepsilon <0$), that facilitate plasmonic resonances --- positioned on the substrate~\cite{huang2003a,eversole2007a}. For example, in the work~\cite{huang2003a}, structures with a $\lambda/260$ size are demonstrated. Several other laser methods for forming nanostructures are considered in~\cite{wang2008a,huang2005nanobump,fardel2010a,huang2011a,wang2007a,wang2008lens-array}.

The opposite effect is also possible --- laser irradiation of the surface through a lattice of spherical particles allows the formation of arrays of nanobumps~\cite{wang2007a,huang2005nanobump} or ``nanohills'' with a height of 10-100 nm, i.e. arrays of nanolenses, see Fig.~\ref{fig:nanobumps}~\cite{wang2008lens-array}. The formation mechanism of the nanobumps is the melting of the substrate substance under the surface of the array of spherical nanoparticles positioned on it and acting as microlenses, with the melt collecting into a nanodrop due to surface tension and its subsequent solidification upon cooling. Surfaces with nano-protrusions may have many different applications. They can be beneficial for the manufacturing of various nanodevices; for extremely high-density information recording, etc.~\cite{huang2005nanobump}.

In conclusion, consider a phenomenon that initially appears to contradict the laws of nature. The statement that light propagates in a straight line through a homogeneous medium seems as undeniable as the fact that water boils at 100 degrees Celsius. However, the adage ``never say never'' holds particularly true in science. While water does boil at 100 degrees Celsius, this only occurs at one atmosphere of pressure. What about the straight light propagation? This is not always the case either.

{In this regard, we note an interesting and important from a practical point of view problem of the formation of a {\it curved photonic jet}, which has received the name {\it photonic hook}. This hook can be used to affect objects in the geometric shadow region --- a gun shooting from around a corner.}

There are various ways to create photonic hooks, a discussion of which can be found in the paper~\cite{Minin2021} and the references therein. As an example, we will cite some results from a recent paper~\cite{hookMinins2024}. It experimentally demonstrated the formation of a photonic hook due to the violation of the axial symmetry of the optical system. Specifically, the hook was created using an optical fiber ends in a truncated cone, whose cross-sectional plane was not parallel to its base; see Fig.~\ref{fig:Hook}. The conical part of the fiber was located vertically above the surface of the polystyrene target. A thermally stabilized laser with a wavelength of 0.671 microns, operating in continuous mode, was used as a radiation source. The exposure time on the target was 6--8 seconds. With these parameter values, the threshold power for producing a crater in polystyrene is slightly less than 10 mW. Note that the crater penetrates the polystyrene in the region of the geometric shadow from the input irradiation spot.

{We also note the work~\cite{red_blood_cells}, which numerically and experimentally proved the possibility of using photonic hooks formed in the same way as in~\cite{hookMinins2024} as cheap and effective optical tweezers. In particular, using such tweezers allows for high-precision movement of individual living cells (erythrocytes), including those situated in areas shielded from the effects of uncurved light beams, i.e., to perform actions that are not feasible using traditional optical methods. }

These results demonstrate the possibilities of using caustics to form structures important for various applications, the characteristic size of which can be significantly smaller than the wavelength of the incident radiation\footnote{{To avoid misunderstandings, it should be emphasized once again that often a significant role in reducing the characteristic scale of the impact of photonic jets on material objects is played by the accompanying nonlinear processes (nonlinear absorption of light, ablation, etc.). Nonlinearity leads to an effective ``steepening'' of the spatial profile of such an impact and, consequently, to a decrease in its transverse dimensions. Discussion of these issues is of independent interest and is not carried out here.}}. They are given only as examples and do not exhaust all possible cases, a detailed discussion of which is beyond the scope of this review.

\section{Phase singularities, optical vortices, superoscillations \label{sec:phase_sing}}
\mbox{}\vspace{-\baselineskip}%

Thus, we have seen that in GO caustics are singular manifolds in which the radiation intensity becomes infinitely large, and one (or both) principal radii of the wavefront curvature vanish. However, describing these singularities within the GO framework is associated with using this approximation beyond its applicability. A more consistent consideration within the framework of wave optics shows that instead of singularities, regions of sharp changes in electromagnetic fields arise, but the fields themselves remain regular --- no infinities arise. In other words, the singularity of caustics is fictitious.
A natural question that can (and should!) be asked in such a situation is: ``Are there singularities in optics whose existence is not associated with exceeding the accuracy of consideration?'' The answer to this question is positive. Moreover, it turned out that the ``zoo'' of such singularities is quite diverse. Within the framework of this review, we have neither the opportunity nor the need {to discuss them all}. Many original and review papers are devoted to such a discussion; see, for example, the already mentioned review by Berry~\cite{berry2023a} and the papers cited there. Therefore, we will limit ourselves to considering the simplest case of phase singularities since we will need a developed approach later for analyzing singularities of the Poynting vector field.

Let us start our consideration again with GO. In the previous sections, we were only interested in ray tracing. We will now consider GO as a certain approximation to solve Maxwell's equations. For simplicity, we will restrict ourselves to considering a monochromatic wave. In the case, when such a wave is plane, all components of the fields $\vec{E}$ and $\vec{H}$ have the form {$A_\nu\exp[i(\vec{k}\vec{r}-\omega t)]$}, where $A_\nu$ denotes any of the three components of any of the specified fields, and $\vec{k}$ is the wave vector {\it in the medium} in which the wave propagates.

Generalizing this expression, we will seek a stationary solution of Maxwell's equations for a monochromatic wave in the form~\cite{LL_Electrodyn,Kravtsov1980} 
\begin{equation}\label{eq:A_Psi}
A_\nu(\vec{r})e^{i\left[\Psi(\vec{r})-\omega t\right]}
\end{equation}
where now $A_\nu(\vec{r})$ --- depends on the coordinates, and the phase gradient\footnote{{In addition to the phase $\Psi$, GO often introduces the quantity $\psi$, called {\it eikonal}, which is related to $\Psi$ by the relation $\Psi = \frac{\omega}{c}\psi$.}} has the meaning of a {\it local\/} wave vector: $\nabla\Psi=\vec{k}(\vec{r})$.

In GO it is assumed that the functions $A_\nu(\vec{r})$ and $\vec{k}(\vec{r})$ are slow, i.e., that they change little on scales of the order of the local radiation wavelength ($\sim 1/|\vec{k}|$), which leads to certain inequalities limiting the applicability range of this approximation~\cite{LL_Electrodyn,Kravtsov1980}. We will not write them out here since we are only interested in the structure of the solution, in which all field components have the same phase that is common to all. This requirement is weaker than the applicability conditions of GO since the solution structure often follows the symmetry of the problem and is not related to the applicability of any approximations.

Now note that a complex quantity can be specified in two equivalent ways --- either in the form of its real and imaginary parts or a modulus and phase. Using Euler's formulas, we uniquely fix the real and imaginary parts by specifying the modulus and phase. In contrast, in the opposite case, we uniquely fix only the modulus by specifying the real and imaginary parts. As for the phase, it is fixed with an accuracy of an arbitrary integer multiple of $2\pi$. However, there is an exception here: when the complex value vanishes. In this case, its modulus is zero, but the phase remains undefined. From what has been said, it follows that the vicinities of points where in the relationship (\ref{eq:A_Psi}) all components of any fields vanish deserves closer attention. Let us consider it in more detail. 

Let, for definiteness, the vanishing field be the field $\vec{E}$. In addition, we will assume that the problem has continuous translational symmetry along the $z$-axis so that the dependence on this coordinate disappears, and $\vec{E}(\vec{r})=\vec{E}(x,y)$. It, however, does not exclude the presence of three nonzero components of $\vec{E}$, so it would be incorrect to call such a problem two-dimensional. Consider the dependence of any one of the components of the field $\vec{E}$ on the coordinates in the neighborhood of the point where $\vec{E}$ = 0. Take, for example, $E_x$. Since this point is not singled out by anything except the fact that there $\vec{E} = 0$, in the general case, the components of $\vec{E}$ at this point must be expanded in a {Taylor} series in powers of the coordinate deviations from this point, and the expansion begins with linear terms. Placing the origin of the coordinate system at the zero point, taking into account that the vector $\vec{E}$ is complex and discarding the higher terms of the expansion, we obtain that
\begin{equation}\label{eq:Ex}
E_x \approx ax+by + i(cx+dy),
\end{equation}
where $a,\;b,\;c$ and $d$ are constants equal to the value of the corresponding derivatives at the zero point. The expansions of the other components of vector $\vec{E}$ look similar. Note that according to Maxwell's {equations} $\vec{H}$ is proportional to the curl of vector $\vec{E}$. Since the curl components are constructed from the first derivatives of the corresponding vector, it follows from this and from the relation (\ref{eq:Ex}) that at the zero point of vector $\vec{E}$ vector $\vec{H}$ is a constant, generally speaking, not equal to zero. A similar statement is true for vector $\vec{E}$ at f zero point of vector $\vec{H}$. Thus, the vanishing of one of the fields does not imply the vanishing of the other.

Moreover, from the nonzero value of the curl of the vector field at a zero point and the Stokes theorem, it follows that for such a field, the curvilinear integral over any closed contour enclosing this point and having small characteristic transverse dimensions is nonzero. It indicates that the zero point is a singularity with a nonzero topological charge. It should be emphasized that this conclusion is based {\it only\/} on the expansion (\ref{eq:Ex}) and Maxwell's equations. It does not use the assumption about the form of the solutions of these equations and, therefore, has much greater generality than the arguments based on the representation of solutions of Maxwell's equations in the form of Eq. (\ref{eq:A_Psi}).

It would seem that we have come to a contradiction. We started with the regularity of the field $\vec{E}(\vec{r})$ at the zero point and concluded that there is a singularity there. To understand this issue, let us turn to the discussed example of solutions {in the form of Eq. (\ref{eq:Ex}). This expression leads to the following coordinate dependence of the phase in the neighborhood of the zero point of vector $\vec{E}$:
\begin{equation}\label{eq:psi_Ex}
\Psi \approx {\rm arctg}\left(\frac{cx+dy}{ax+by}\right),
\end{equation}

In this case, due to the assumption of identity of $\Psi(\vec{r})$ for all field components, the expression (\ref{eq:psi_Ex}) is also valid for the phases of the components $E_{y,z}$. Moreover, it is valid literally, i.e. {\it with the same values\/} of the constants $a,\;b,\;c,\;d$.

Then, the calculation of $\nabla\Psi$ gives the value of the local wave number equal to
\begin{equation}\label{eq:k_local}
k \approx \left|\frac{(b c-a d)\sqrt{x^2+y^2}}{x^2 \left(a^2+c^2\right)+2 x y (a b+c d)+y^2 \left(b^2+d^2\right)}\right|
\end{equation}

In the expression {(\ref{eq:k_local})}, the numerator has the first order of smallness in $x$ and $y$, and the denominator has the second. Therefore, as the observation point approaches the coordinate system origin, when \mbox{$\sqrt{x^2+y^2}\rightarrow 0$}, the value of $k$ diverges. The divergence of the local wave number means the presence of spatial oscillations in this field, with a period tending to zero as the origin is approached. On the other hand, we started with the assumption that the field in the vicinity of the origin can be represented as an expansion (\ref{eq:Ex}), which is a linear function of the coordinates and generally {\it does not contain any oscillations!}. Perhaps the oscillations indicated are a mathematical artifact that has nothing to do with the actual structure of the field. Is it so, indeed?

Let us turn to the expression (\ref{eq:psi_Ex}) to understand what is going on here. It is easy to see that the family of {lines} of constant phase $\Psi(x,y)=\Psi_0=const$ in this expression is a fan of straight lines emanating from the coordinate center and covering the entire space. The gradient of a scalar function is oriented perpendicular to the lines of its constant value. In our case, it follows that the vector $\vec{k}$ is perpendicular to these lines and makes a complete revolution when going around the origin along a closed curve. In this case, the phase of the vector $\vec{E}$ acquires a phase shift of $\pm 2\pi$ (the specific sign is determined by the relationship between the coefficients $a,\;b,\;c,\;d$). The smaller the scale of the contour, the shorter the path required to obtain a complete revolution of the vector $\vec{k}$, the faster the phase changes, i.e., the greater its gradient, and, hence the larger the modulus $|\vec{k}|$. These are the spatial oscillations that we obtained above.

{We emphasize that so far, all the considerations concern only the phase of the field of vector $\vec{E}$ and its local wave vector. The phase of the light wave can be measured (at least in principle) from the interference pattern of a reference beam, the phase of which is known, with our singular wave field. Nevertheless, we would like to understand how the vector $\vec{E}$ itself behaves. To this end, we note that, due to the topological charge conservation, the conclusion about the change in the phase of $\Psi$ by $\pm 2\pi$ when going around a singularity along a closed contour remains valid not only for a contour lying in the immediate vicinity of the singularity but also for a contour {\it of any size\/} as long as it does not encompass other singularities. However, as discussed above, at a sufficient distance from the singularity, the applicability of GO is restored, even if GO is inapplicable in its immediate vicinity. The applicability of GO guarantees that in the expression (\ref{eq:A_Psi}), the amplitudes $A_\nu(x,y)$ are slow functions of the coordinates. In this case, in the equation ${\rm div}\/\vec{E}=0$ in the leading approximation {${\rm div}\/\vec{E}\approx\vec{E}\nabla\Psi=\vec{E}\vec{k}$,} i.e. the vector $\vec{E}$ is orthogonal to the vector $\vec{k}$, and the electromagnetic wave is completely transverse. Then, it immediately follows that if the passage along a closed contour leads to the rotation of vector $\vec{k}$ by $\pm 2\pi$, the same happens with vector $\vec{E}$. Repeating the reasoning about the topological charge conservation in reverse order, we conclude that this property of vector $\vec{E}$ is preserved in a small neighborhood of the singular point, even if the GO approximation is violated in this region, and the electromagnetic wave acquires longitudinal components. Such a singularity is called {\it an optical vortex}.}

Thus, the phase singularity we have considered is related to the vector nature of the fields $\vec{E}$ and $\vec{H}$ and manifests itself in a rapid ``rotation'' of the corresponding vector around the singular point, and not at all in rapid {spatial} oscillations of each of its components. The latter only changes signs when passing through the singularity.

{It may seem that from expression (\ref{eq:A_Psi}) it follows that since the phase of all components of the vector $\vec{E}$ is the same, and their ratio does not depend on $\Psi$. Then, a change in $\Psi$ varies only the ratio between the real and imaginary parts of each component, but not the orientation of the vector $\vec{E}$ in space. It contradicts the statement just made. To understand where the error is hidden, we note the following. First, we need to define {{\it what\/}} we mean by the direction of a complex vector, since the directions of ${\rm Re}\,\vec{E}$ and ${\rm Im}\,\vec{E}$ are, generally speaking, different. Since the complexity of the fields included in Maxwell's equations is introduced solely for the convenience of calculations, and the actual physical fields are real, we take the direction of the real part of vector $\vec{E}$ for the direction of the entire vector. Secondly, due to the complexity of the amplitudes $A_\nu$ in expression (\ref{eq:A_Psi}), in the vicinity of the singularity, the components of vector $\vec{E}$ can be rewritten as follows:
\begin{equation}\label{eq:Psi_0}
E_{\nu}=|E_\nu|\exp[i(\Psi(x,y,z)+\Psi_\nu^{(0)})];\;\;\Psi_\nu^{(0)}=const;\;\; \nu = x,\;y,\;z.
\end{equation}
Here the function $\Psi(x,y,z)$ is the same for all components, but the constants $\Psi_\nu^{(0)}$ differ from each other\footnote{{Note that the question of separating the contributions to the total phase from the function $\Psi(x,y,z)$ and $\Psi_\nu^{(0)}$ is not entirely trivial. In fact, it comes down to the underlying GO representation of fields in the form~(\ref{eq:A_Psi}), for more details see, for example,~\cite{Kravtsov1980,kravtsov1993a}.}}. It violates the proportional change of various components of the vector ${\rm Re\,}\vec{E}$ when $\Psi$ changes and ensures its rotation around the singularity when an observation point goes around it.} 

Note that {abrupt changes} of the phase by $2\pi$ when going around the point at which the modulus of the complex function vanishes are well-known in other areas of physics, where they also play a significant role. The spectrum of such areas extends from superconductivity to pattern formation in macroscopic nonequilibrium systems~\cite{Ivlev_UFN1984,JETP_PhaseSlip_Golberg,golberg1989_JLowTemp,tribelsky1995_PRE_PhaseSlip} --- an example of the fact that physics is unified, and phenomena that are distant from each other, seemingly having nothing in common, are described by the same approaches.

{Let us make another important remark. {The discussed optical vortex formation at a zero point of one of the electromagnetic fields is essentially connected with the possibility of describing all components of these fields by a single phase. If such a description is impossible, then the above arguments are not applicable to such fields. For example, a standing wave is formed when two monochromatic plane linearly polarized waves with the same polarization and equal amplitude interfere, propagating towards each other. At its nodes, the corresponding fields vanish. However, this does not lead to the formation of optical vortices, since in this case, the field is represented as a sum of terms for which the phases $\Psi$ differ in sign, and its description in the form of Eq. (\ref{eq:A_Psi}) is impossible. }

As for the discussed spatial oscillations with wave numbers significantly exceeding those included in the Fourier spectrum of the corresponding field, they are called {\it superoscillations\/}~\cite{berry2006a,rogers2012a,berry2019roadmap,aharonov2017a} and are found in a wide range of optical phenomena, see, for example~\cite{berry2019roadmap} and the references therein. We especially emphasize that superoscillations are not only a beautiful phenomenon --- they are essential in practice, for example, in subwavelength microscopy~\cite{wang2016a,wang2011a,astratov2023a}.

Let us now consider the influence of phase singularities on the manifestation of the diffraction limit. So far we have discussed the diffraction limit in connection with focusing of optical radiation, which tacitly implies an increase in the intensity of light. At phase singularities, at least one of the fields (either $\vec{E}$ or $\vec{H}$) vanishes, and the phase of the vanised field is undefined. In addition, since the Poynting vector is proportional to the vector product of the fields $\vec{E}$ and $\vec{H}$, vanishing either of these fields turns the Poynting vector (i.e., the radiation intensity) to zero. How does all this affect the diffraction limit?

Again, we will use the Heisenberg uncertainty principle, but now we will apply it to the energy of a light wave: $\Delta U\Delta t \sim \hbar$. As for the energy, it is obvious that $U=N\hbar\omega$, where $N$ is the number of photons. $\Delta t$, which is included in the uncertainty relation, is the lifetime of the given quantum state. Taking into account the time factor, the expression for the total phase in the GO approximation has the form {$\Psi-\omega t$}. Therefore, the lifetime can be estimated as $\Psi/\omega$. Putting everything together, we obtain that, in the case under consideration, the uncertainty principle takes the form $\Delta N\Delta \Psi \sim 1$. Since at the point of the phase singularity the phase is not defined, i.e., it has infinite uncertainty, it means that the number of photons at this point is zero. Comparing this statement with the above reasoning about the presence of a phase singularity at the point where the amplitude of one of the electromagnetic fields vanishes, we come to the conclusion that the converse statement is also true: the presence of a phase singularity leads to the vanishing of the radiation intensity at this point.

Next, as is usually done in quantum mechanics, applying the estimate $\Delta N \sim N$, and $\Delta \Psi \sim \Psi$, we obtain that $\Psi\sim1/N$. This yields the following estimate of the local wave number: {$k\equiv |\nabla \Psi|\sim |\nabla N|/N^2\sim 1/(LN)$, where $L$ is the characteristic spatial scale of the profile $N(\vec{r})$.} This estimate explains the divergence of $k$ in the ``dark light'' regions, where \mbox{$N\rightarrow 0$.}

{We also note that in the vicinity of the phase singularity, the characteristic spatial scale of the field is determined by $1/k$. The divergence of $k$ means that when approaching the singular point, this scale tends to zero, which, as follows from the previous paragraph, {\it does not violate the diffraction limit.} This allows one to create optical vortices of substantially subwavelength sizes~\cite{wang2019a,berry2001knotted,leach2004knotted,wang2004a,bashevoy2005a,tribelsky2006a,Lukyanchuk2007,s2008a,luk2013,kuznetsov2016a}.}

The natural questions that arise from the above discussion are: ``How does all this affect the energy flow (the Poynting vector field) in the vicinity of such singularities, and are there singularities in this field that are not associated with the vanishing of the $\vec{E}$ or $\vec{H}$ fields?'' The next section of the review is devoted to clarifying the answers to these questions.

\section{Poynting vector field singularities\label{sec:Poynting}}
\mbox{}\vspace{-\baselineskip}
\subsection*{Introductory remarks\label{subsec:Pointing_Introduc}}
The Poynting vector field {describes} the energy transfer by an electromagnetic wave, and the divergence of this field determines the density of electromagnetic energy dissipation in matter. For this reason, knowledge of the structure of the Poynting vector field is of paramount importance, both from an academic point of view and from the point of view of numerous applications, especially {related to} medicine, biology, as well as information processing and recording, where time- and space-controlled energy release plays a fundamental role. {Some of them (laser surface cleaning, creation of metamaterials with unique properties that do not exist in natural analogs, etc.) have already been mentioned above. Among others, we can name nanosurgery of living cells and individual chromosomes~\cite{nanosurg,berns2020laser}. Precise control of the electromagnetic energy density on subwavelength scales is important in manufacturing modern optoelectronic micro- and nanodevices~\cite{Gaweda2009}. Other examples of using Poynting vector field structuring in various nanotechnologies can be found, for example, in the reference book~\cite{Bhushan2017}. Note also that the type and position of the Poynting vector field singularities affect the localization of the point of application of the angular momentum of ponderomotive forces~\cite{Mokhun2008}, which can be used to create optical traps and tweezers. This list can easily be continued.}

{Note also that the Poynting vector field is not included in Maxwell's equations and is not directly determined by them. It {\it is calculated\/} from the known profile of fields $\vec{E}$ and $\vec{H}$. Though direct measurement of the Poynting vector field, although is possible, the corresponding experiments} are a difficult task~\cite{Mokhun2012}. In such a case, theoretical studies come to the fore, particularly the study of singular points of the Poynting vector field, since it is the type and spatial localization of singularities of the vector field that largely determine its global topological structure. This section is devoted to a discussion of these issues. However, before doing that, we will make several important general remarks.

We will use the standard definition of the Poynting vector as a real value, which for a monochromatic wave, after averaging throughout the field temporal oscillations, is given by the relation
\begin{equation}\label{eq:S}
\vec{S}=\frac{c}{16\pi}({[\vec{E}^*\vec{H}]} +{[\vec{E}\vec{H}^*]}),
\end{equation}
where the asterisk denotes complex conjugation. {In some cases, it also makes sense to introduce the complex Poynting vector $\hat{\vec{S}} = \frac{c}{8\pi}[\vec{E}^*\vec{H}]$, the imaginary part of which {describes the oscillating flow} of the stored energy~\cite{Jackson1998}. This quantity plays an important role in some problems of  light --- matter interaction~\cite{bliokh2014magnetoelectric,Bliokh2014,Bekshaev2015,Xu2019,Khonina2021,Tang2010,Lininger2022}. In most cases, the results discussed below, obtained for a real $\vec{S}$, can easily be generalized to the case of a \mbox{complex $\hat{\vec{S}}$.}

The Poynting vector field will be described by lines of force, which by analogy with hydrodynamics we will also call {\it streamlines.} By definition of lines of force of a vector field, at each point of such a line, the vector $\vec{S}$ must be tangent to it. It is convenient to specify the equations of lines of force in parametric form: $\vec{r}=\vec{r(t)}$. Then, each such line can be considered as a trajectory of motion in the space of a virtual material point. In this case, $t$ will play the role of ``time''.

We put the word ``time'' in quotation marks to emphasize that this is just a parameter that has no relation to the actual dynamics of energy flow transfer along streamlines. However, the interpretation of a streamline as a trajectory turns out to be very useful since it immediately allows us to write an equation describing the shape of this trajectory. Indeed, when a material point moves along a trajectory, its velocity at each point is directed along a tangent to this trajectory. It follows immediately that the vector $\vec{S}$ is proportional to $d\vec{r}/dt$. On the other hand, since the parameter $t$ is not yet defined, the proportionality coefficient in this relationship can always be converted to unity by appropriately scaling $t$. As a result, we obtain that the equation describing the streamlines of the Poynting vector field reads as follows.
\begin{equation}\label{eq:S_lines}
\frac{d\vec{r}}{dt}=\vec{S}(\vec{r}).
\end{equation}

In this case, the usual Poincar\'e classification applies to the singularities of the Poynting vector field, in which a singularity is an intersection point of the null-isoclines $S_{x,y,z}(\vec{r}) = 0$, and its type is determined by the roots of the characteristic equation~\cite{Arnold_ODE}; see also below. Such a classification is well known; see, for example, the work~\cite{Novitsky2009}. In this case, according to formula (\ref{eq:S}), the Poynting vector field inherits all the singularities of the fields $\vec{E}$ and $\vec{H}$. In particular, the zeros of each of these fields are simultaneously zeros of the Poynting vector field, i.e., they satisfy the condition $S_{x,y,z}(\vec{r}) = 0$, and therefore, are singularities of the streamlines of this field. According to the established terminology, such singularities will be called {\it field-type singularities}, in particular, {\it E}-type for singularities associated with the vanishing of the electric field, and {\it H}-type when the magnetic field vanishes.

However, the case is not limited to this. As is known, in a standing wave, the energy flow averaged throughout the field temporal oscillations vanishes not only at the wave nodes but also in regions where neither of fields $\vec{E}$ and $\vec{H}$ equal zero. Accordingly, singularities of field $\vec{S}$ can form at points where neither of fields $\vec{E}$ and $\vec{H}$ vanish. Such singularities will be called {\it polarization-type singularities}.

At the same time, despite the intensive study of the features of the Poynting vector field; see, for example, works~\cite{wang2004a,bashevoy2005a,tribelsky2006a,Lukyanchuk2007,Lukyanchuk2007a,Mokhun2012,CanosValero2021,TribelMirosh_UFN_2022,Luk2006_wire}
and the references therein; insufficient attention has so far been paid to the fact that this field must satisfy the energy conservation law and the constraints imposed by the symmetry of the problem. These conditions lead to a significant decrease in the ``number of degrees of freedom'' of the problem of the Poynting vector field singularities compared to the situation valid for a vector field that does not have such constraints. Recent studies ~\cite{Tribelsky2022,Tribelsky2022a,TribelskyJETPL,tribelsky2024_LPR}, have made some progress in addressing these issues, as discussed briefly below. However, the task is still far from complete.

Finally, let us delve into the intriguing question of the characteristic spatial scale of singularities. As the characteristic scale of the problem tends to zero near the singular point (as discussed in the context of superoscillations), it becomes logical, as was previously highlighted in the Introduction, to focus on the upper boundary that defines the region where the field structure is predominantly influenced by the given singularity. This leads to a qualitative difference between traditional singularities found in speckle structures~\cite{Dainty1984}, paraxial beams~\cite{Angelsky2022}, etc. and the singularities emerging in resonant light scattering by nanoobjects~\cite{tribelsky2006a}. IIn the former case, singularities typically emerge due to the properties of the incident laser beam, with their upper scale being at least equal to—and often significantly greater than—the wavelength of the incoming radiation. In contrast, in the latter case, singularities arise spontaneously, even during the scattering of a linearly polarized plane wave that initially has no singled out points at all. Here, the characteristic scale of the region where the field structure is primarily influenced by a singularity is significantly smaller than the wavelength of the incident radiation.

Note also that {in many cases}, the spatial configuration of the incident electromagnetic radiation is limited from below by scales of the order of the wavelength of this radiation. At {smaller scales}, such a wave can be considered locally as a plane wave. It significantly expands the applicability range of the results of studying the problem of subwavelength singularities in the scattering of a plane wave.

The physical basis for the existence of subwavelength singularities in the scattering of light by nanoparticles has been previously discussed: in such scattering, the wavelength of the radiation ceases to play the role of the characteristic scale of the problem. The only characteristic scale in this case is the size of the scattering object, which manifests itself in the structure of the near field and, in particular, in the upper scale of singularities. It's important to note that such subwavelength singularities are not just theoretical constructs, but they have significant practical applications. They allow us to use the emerging structure of the electromagnetic field to influence material objects with ultra-high spatial resolution, and to study various quantum effects~\cite{wang2011a,wang2016a,krivitsky2013locomotion,huang2018planar,astratov2023a,barnett2013momentum,berry2004quantumCore,barnett2008quantumCore}.

\subsection*{Symmetry, energy conservation law, invariant planes and two-dimensional structures\label{subsec:Symmetry}}

The following description of the Poynting vector field singularities is based on the assumption that we have already used earlier in discussing the singularities of the phase of $\vec{E}$ and $\vec{H}$ fields: at a singular point, the field components vanish, yet remain regular {functions} expandable in a {Taylor} series in small deviations of the coordinates from this point. Everything else is a matter of technique. Choosing the singularity as the origin of the local coordinate system; expanding $S_{x,y,z}(x,y,z)$ in such series, restricting ourselves to linear terms; taking into account that, due to the symmetry of the problem (if any), some expansion coefficients can vanish identically, while the remaining ones can have {certain properties} concerning the change of the sign of the coordinates; supplementing this with the law of conservation of energy, according to which $-$div$\vec{S} = Q$, where $Q$ is the density of dissipated electromagnetic energy, it is possible not only to carry out a complete classification of singular points but also to construct a complete bifurcation picture of their birth, motion and annihilation, arising when changing various parameters of the problem. A detailed discussion of these issues is beyond the scope of this review, so below, we will present only the principal results and several characteristic examples, referring readers interested in details to the original works~\cite{Tribelsky2022,Tribelsky2022a,TribelskyJETPL,tribelsky2024_LPR}.

A typical example of this kind of problem is the structure of the field during scattering of a plane linearly polarized wave by a sphere, which was studied in the works~\cite{wang2004a,bashevoy2005a,tribelsky2006a}. Fig.~\ref{fig:Sphere_lines} shows such a structure in the equatorial plane of the scattering sphere. For a given value of the size parameter $q=kR$, the chosen value of the permittivity corresponds to the neighborhood of the electric dipole resonance. The exact resonance is achieved at $\varepsilon = -2.2229...$

To determine the electromagnetic type of singularities, the Poynting vector streamlines are superimposed on the field profiles $|\vec{E}|^2$ and $|\vec{H}|^2$, plotted in a logarithmic scale. It is seen that the foci correspond to the field singularities of the $H$-type, and the saddles correspond to the polarization singularities. It is a general property of the problem and has a simple explanation. Indeed, polarization singularities are associated with the formation of standing waves, which are two traveling waves of equal amplitude propagating toward each other. It implies the presence of preferred directions in the problem. For foci, such directions do not exist; therefore, polarization-type singularities are impossible for them. On the contrary, saddles have distinguished directions, coinciding with their separatrices whiskers.

Note, however, that, strictly speaking, this statement applies only to two-dimensional field patterns. In three dimensions, a two-dimensional focus can have a third root of the characteristic equation that is purely real (a {saddle-focus} singularity). In this case, forming a standing wave is possible in the direction determined by the eigenvector corresponding to this root. Some aspects of field singularities in three-dimensional space will be discussed below, but in general, this area has not yet been thoroughly studied. 

It is important to emphasize that, as already noted, in the problem under consideration, the characteristic size of the field structures is much smaller than the radiation wavelength. Of course, this also applies to the discussed standing waves, which are realized only in an essentially subwavelength neighborhood of the singular points. %This underscores the need for further research and exploration in this area, which is crucial for advancing our understanding of field singularities.

Further, in the situation depicted in Fig.~\ref{fig:Sphere_lines}, it follows from the symmetry of the problem that the plane of the figure is an invariant plane --- the streamlines lying in this plane do not go out of it. It means that in the plane of the figure, field $\vec{E}$ has only two components different from identical zero: $E_{x,z}$, which depend only on two variables, $x$ and $z$. In this case, field $\vec{H}$, which according to Maxwell's equations is proportional to the curl of $\vec{E}$, has a single component different from identical zero, oriented perpendicular to the plane of the figure and depending on the same two variables.

Now it is important to note that in (\ref{eq:S}), $\vec{E}$ and $\vec{H}$} are complex quantities. Consequently, for $\vec{S}$ to vanish at the field singularity, both the real and imaginary components of the corresponding fields must also vanish. In this case, the coordinates of the field singularity on the $xz$ plane are determined by solving the corresponding system of equations. The requirement that the field $\vec{E}$ be zero leads to four equations: Re\/$E_{x,z}(x,z)=0$, \mbox{Im\/$E_{x,z}(x,z)=0$,} for two coordinates $x$ and $z$. In general, such a system is overdetermined and has no solutions. The exceptions are situations when one of the components of the field $\vec{E}$ identically vanishes due to the symmetry of the problem. For example, in the case shown in Fig.~\ref{fig:Sphere_lines}, due to the mirror symmetry of the structure relative to the $z$-axis, for any point belonging to this axis, $E_x$ is identically equal to zero. However, in this case, the $x$ coordinate is fixed by the condition $x=0$. We obtain two equations Re\/$E_{z}(0,z)=0$, \mbox{Im\/$E_{z}(0,z)=0$} for the only remaining coordinate $z$, which again leads to an overdetermined system of equations that has no solutions.

On the other hand, if we require the vanishing at the field singularity of field $\vec{H}(x,z)$, which has a single component different from identical zero, no complications of this kind arise. The requirement $\vec{H}=0$ leads to the two equations for the two variables: {Re\/$H_y(x,z)=0$, Im\/$H_y(x,z)=0$,} the solution of which determines the positions of the singular points.

The above reasoning is quite general. It is valid in any case when the symmetry of the problem leads to the emergence of an invariant plane for the streamlines. Then, if vector $\vec{E}$ lies in the invariant plane, field $\vec{H}$ vanishes in a field singularity. If the invariant plane coincides with the plane where vector $\vec{H}$ lies, field $\vec{E}$ turns to zero at a filed singular point.

Let us discuss some more interesting topological features of the Poynting vector field depicted in Fig.~\ref{fig:Sphere_lines}(a). The figure shows that the streamlines enclosed within the separatrix emerging from the saddle, designated by number 3 in the figure, and enclosing the scattering sphere are not connected with the surrounding space. It means that the energy of the electromagnetic field circulates along these lines, being locked in a finite region of space. Since the field there was absent before the incident light came, it raises the question of how it appeared. The only possible answer to this question is that at the initial stage of irradiation when the scattering process was non-stationary, the topological structure of the field was different. Numerical calculations confirm this conclusion. They show that at the non-stationary stage of the scattering process, the topological structure of the scattered field and the one inside the scattering particle, undergoes qualitative changes associated with the birth and spatial movement of various singularities~\cite{svyakhovskiy2018transient}. However, research into this problem is in its infancy; many essential aspects have not yet been revealed, and the relevant questions have not even been formulated.

Another essential remark concerns the foci indicated in Fig.~\ref{fig:Sphere_lines}(a) by numbers 4 and 5. It is evident that both foci are unstable --- the srteamlines emerge from them, i.e., energy continuously flows out of these singularities. {On the other hand, the presented picture of the streamlines is stationary. Therefore, by virtue of the law of conservation of energy, the energy flowing out of the foci must be compensated by its inflow in some other directions. These ``other'' directions must be transversal to the invariant plane. That is precisely what happens. Due to the three-dimensionality of the problem, in addition to the two eigenvectors lying in the invariant plane, each of the singularities shown in Fig.~\ref{fig:Sphere_lines}(a) has third eigenvector perpendicular to this plane.} %It would seem that this fact is in blatant contradiction with the law of conservation of energy. In fact, there is no contradiction. The figure shows the structure of the field in the invariant plane, i.e., on a two-dimensional manifold. In reality, the problem is three-dimensional. Therefore, in addition to the two eigenvectors lying in the invariant plane, all the singular points shown in the figure have a third one --- orthogonal to this plane.

The analysis carried out taking into account the restrictions mentioned above imposed by symmetry, and the law of conservation of energy shows that the root of the characteristic equation corresponding to this third vector is always purely real. In this case, for an unstable two-dimensional focus, the corresponding direction in the perpendicular plane is stable, and for a stable one, it is unstable. Therefore, the energy flow emerging from foci 4 and 5 in directions close to the invariant plane is compensated by the inflow of energy along directions close to the normal to this plane. %

If the problem were two-dimensional, foci in a non-dissipative medium would be prohibited because it would contradict the energy conservation law. This situation occurs when an infinite circular cylinder is irradiated by a linearly polarized plane wave whose wave vector is perpendicular to the cylinder axis (normal incidence), and the plane of polarization is either perpendicular to this axis (the so-called TE polarization), or the axis of the cylinder lies in the plane of polarization (TM polarization). In these cases, in the absence of dissipation, only singular points of the center or saddle type are possible~\cite{Tribelsky2022}.

In the case of light scattering by a particle whose material absorbs the incident {radiation}, the energy flow along the field lines of the Poynting vector is accompanied by its dissipation. Therefore, the existence of closed loops of field lines that at least partially pass inside the scattering particle becomes impossible. %, as contradicting the law of conservation of energy.
{Indeed, let us assume that a closed loop does exist. Such a streamline has no connection with the outside world, and no energy enters it. Whereas, dissipation occurs, and electromagnetic energy continuously turns into heat. In the absence of external compensation for such losses, this cannot be the case}.

In particular, all singularities of the center type situated inside the particle turn into foci. We emphasize that we are talking only about singularities located {\it inside\/} the scattering particle. As noted in the Introduction, the environment surrounding the particle is assumed to be non-dissipative, therefore all the above-mentioned restrictions, valid for non-dissipative substances, remain in force there.

Note also that the presence of dissipation violates the symmetry between fields $\vec{E}$ and $\vec{H}$ that exists in non-dissipative media, and as a consequence, between the field singularities of the $E$- and $H$-types. The point is that when electromagnetic radiation is absorbed by a medium, the density of the dissipated power is proportional to the sum $\varepsilon''|\vec{E}|^2 + \mu''|\vec{H}|^2$~\cite{LL_Electrodyn}, where two primes denote the imaginary parts of the corresponding quantities. However, at optical frequencies, the magnetic permeability $\mu$ is purely real and equal to unity~\cite{LL_Electrodyn}. Therefore, the term with the magnetic field vanishes, and the dissipative processes become associated only with the electric field. As a result, if in $H$-type singularities, at the center of which the electric field is not equal to zero, the presence of dissipation transforms the singular points of the center type into foci already in the linear approximation with respect to the deviation of coordinates from the singularity point, for $E$-type singularities, at the singular point itself the density of the dissipated power is equal to zero, and in the linear approximation the singularity type remains the center. It transforms into a focus only taking into account quadratic corrections. Such foci are called slow. They have a significantly smaller helix pitch and a faster decrease in this pitch as the field line approaches the singular point than is inherent in a traditional focus~\cite{Tribelsky2022a}.

\subsection*{Soft symmetry breaking, three-dimensional structures, bifurcations\label{subsec:Symmetry_violation}}
So, we have seen that the law of energy conservation and the symmetry of the problem play a significant role in forming singularities of the Poynting vector field and impose certain restrictions on their type and properties. A natural question arises in this connection: "What happens if some of these restrictions are lifted?" Of course, we cannot cancel the law of conservation of energy. Not so with symmetry. There are many ways to break it. In this case, the most significant interest is presented by "soft" symmetry breaking, when the problem has a continuously changing parameter, the value of which determines the degree of symmetry breaking. For this purpose, in the work~\cite{tribelsky2024_LPR} the structure of the field was investigated during the scattering of a linearly polarized plane wave by an infinite cylinder of circular cross-section when the wave vector of the incident wave was still perpendicular to the axis of the cylinder, but the plane of polarization could be oriented arbitrarily, so that vector $\vec{E}$ formed an arbitrary angle $\alpha$ with the axis of the cylinder, see Fig.~\ref{fig:Oblique}.

This problem formulation preserved translational symmetry along the cylinder's axis. In contrast, mirror symmetry relative to the plane perpendicular to the cylinder's axis was controllably violated by changing the angle $\alpha$. Due to translational symmetry, the components of the Poynting vector cannot {depend} on the coordinate $z$. As for the dependence on $x$ and $y$, it can be shown~\cite{tribelsky2024_LPR} that {in such a problem} they satisfy the relations:
\begin{equation}
(S_x(x,y),S_y(x,y),S_z(x,y)) =
(S_x(x,-y),-S_y(x,-y),-S_z(x,-y)). \label{eq:S_y->-y}
\end{equation}
For an arbitrary value of $\alpha$, the scattering problem admits an exact analytical solution~\cite{bohren2004a}. It significantly simplifies the study of the arising singularities. Such a study shows that a change in the problem's parameters ($\alpha,\;\varepsilon,\;q$) leads to a complex scenario of successive bifurcations of the creation and annihilation of singular points. In the vicinity of the bifurcation points, one can develop a phenomenological theory grounded solely in the constraints set by condition (\ref{eq:S_y->-y})and the energy conservation law. This theory governs the coefficients of the expansion of the Poynting vector components in small deviations of the coordinates from specific characteristic points. The predictions of the bifurcation picture following from the theory ideally coincide with the results obtained in the analysis of the exact solution of the problem. However, in our opinion, a detailed discussion of the bifurcation picture is of interest only to narrow specialists. We do not present it here, referring the interested reader to the original work~\cite{tribelsky2024_LPR}.

We note only one significant difference between the problem with broken symmetry and its more symmetric analog. In the former case, all streamlines, except the lines belonging to the invariant plane $y=0$ (see Fig. \ref{fig:Oblique}), become three-dimensional. On the other hand, the position of singularities should be determined by the intersection points of three surfaces $S_{x,y,z}=0$. As a result, to determine the coordinates of the singularities, we obtain {\it three\/} equations for {\it two\/} variables $x$ and $y$, see (\ref{eq:S_y->-y}). Again, we obtain an overdetermined system that has no solutions. The exceptions are cases when, on some manifold, components of the Poynting vector vanish identically due to the symmetry of the problem. From the continuity of the coordinate dependence of the components of the Poynting vector field and the condition (\ref{eq:S_y->-y}) it follows that \mbox{$S_y(x,0)=S_z(x,0)=0$}. In other words, the specified manifold is the plane $y=0$, in which the two components of the Poynting vector are identically equal to zero. Then, the condition specifying the position of the singularity is reduced to a single equation $S_x(x,0)=0$, which may or may not have a solution --- this is determined by the specific type of dependence $S_x(x,0)$. One way or another, but now the singularities of the Poynting vector field can only be in the plane $y=0$. Moreover, due to continuous translational symmetry along the $z$-axis, the singularities (if any) are not separate points, but continuous lines parallel to the $z$-axis.

In addition to the three-dimensional Poynting vector field pattern discussed above, it is helpful to consider its two-dimensional projection onto a plane perpendicular to the cylinder axis. Such a two-dimensional picture may also contain singular points. However, it should be noted that if these points do not belong to the $x$-axis, they are regular in three-dimensional space, since they have a non-zero value of the $S_z$ component. In the work~\cite{tribelsky2024_LPR}, such singularities were called {\it false}.

Fig. \ref{fig:2d-3D} illustrates the above. Note that the spiral streamline shown in Fig.~\ref{fig:2d-3D}(b) does not have translational symmetry along the $z$-axis, which seemingly contradicts the initial assumptions. In fact, there is no contradiction. The translational symmetry of the problem guarantees that $\vec{S}(x,y,z+a)=\vec{S}(x,y,z)$ where $a$ is an arbitrary constant of the appropriate dimension. It does not necessarily mean that any streamline must be transformed into itself by such a transformation. To achieve translational invariance, it is sufficient for a translation to transform a given streamline into another streamline that retains the same shape as that for the original. The streamline depicted in Fig.~\ref{fig:2d-3D}(b) meets this criterion. 

{Note that such three-dimensional spirals, winding on some (not necessarily straight) line or unwinding from it, are typical structures of the Poynting vector field. They are formed not only during resonant scattering of an electromagnetic wave by a cylinder but also during scattering by spherical particles. In the latter case, their appearance is usually associated with the formation of a singular point of the saddle-focus type, and the line around which the spiral is wound (unwound) is the saddle whisker, see, for example, {~\cite{wang2004a,2006_OptJ,Lukyanchuk2007,Lukyanchuk2007a,luk2013,Luk2017PhilTran}.}

\section*{Conclusion\label{sec:Concl}}
\mbox{}\vspace{-\baselineskip}

To summarize, we have considered various types of light beam singularities, sequentially descending from micron scales and GO to subwavelength optics. We have discussed that ray tracing can lead to the formation of caustics. In the framework of GO, they are singular formations on which the radiation intensity becomes infinite, and at least one of the radii of curvature of the wave front turns to zero. Increasing the accuracy of considering caustics when moving to wave optics shows that geometric caustics are blurred in the regions of high but finite radiation intensity, having a finite extent in all three spatial coordinates and their own internal structure.

Unlike caustics, phase singularities correspond to the vanishing of field $\vec{E}$ or/and $\vec{H}$, and therefore, the radiation intensity. In this case, the fields themselves can remain monotonic, expandable in a Taylor series of functions of coordinates, but their local wave number (the modulus of the phase gradient) diverges at the singularity point. For this reason, phase singularities demonstrate superoscillations associated with the rotation of the field vector by $\pm 2\pi$ when going around the singularity along a closed contour.
Similar singularities are demonstrated by the Poynting vector field, but in this case a new type of singularities can form, caused by local excitation of standing waves, when neither of fields $\vec{E}$ and $\vec{H}$ vanishes at the singular point.

The symmetry of the problem plays an important role in the formation of the properties of singularities. In the case of singularities of the Poynting vector field, the law of conservation of energy imposes additional restrictions on these properties.

The use of singularities and their properties helps to overcome the diffraction limit and control electromagnetic radiation on subwavelength scales, which is important for a wide range of applications, some examples of which were given in various sections of this review. We hope that our review clarifies a number of important issues of subwavelength optics and stimulates further research in this promising and rapidly developing area.

\bigskip

The authors are grateful to Yu.S. Kivshar, I.V. Minin, O.V. Minin, A.V. Novitsky and O.I. Surov for useful discussions and valuable comments. Our special thanks to N.D.~Arnold for a detailed discussion of issues devoted to geometrical optics and caustics, as well as for his valuable comments, which helped us to significantly improve the presentation of these issues in the review.

\newpage
\begin{figure}[tbh!]
\centering
\includegraphics[width=\textwidth]{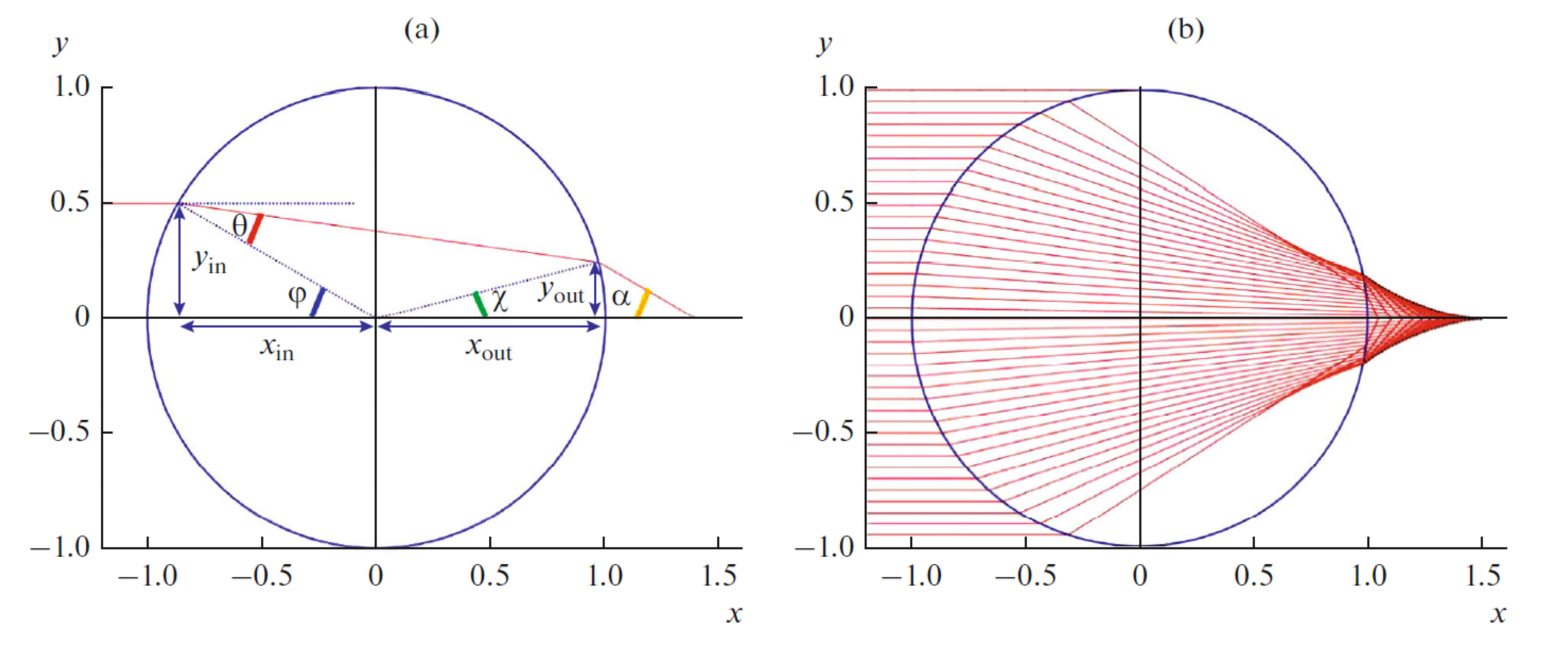}
\caption{(a) Ray tracing in the equatorial plane of a sphere of radius $R$ irradiated by a parallel light beam in a vacuum. The refractive index of the sphere is $n = 1.5$. The $x$ and $y$ coordinates are normalized to $R$. The angles of incidence ($\varphi$) and refraction ($\theta$) inside the sphere are related by Snell's law: $\sin\varphi = n \sin \theta$. The ray enters the particle at the point with coordinates $y_{\rm in} = {\rm tg}\, \varphi$ and $x_{\rm in} = - \cos\varphi$. The angles $\chi$ and $\alpha$ are defined by the formulas $\chi = 2\theta \,- \,\varphi$ and $\alpha = 2(\varphi - \theta)$. Two nearby rays, $y_{\rm c}$ and $y_{\rm cc}$ (corresponding to the angles $\varphi$ and $\varphi+\delta\varphi$), intersect at the caustic point at the exit of the sphere after the second refraction, which gives the equation for the caustic (\ref{eq:cusp}). (b) The shape of the caustic (the curve bounding the rays leaving the sphere) is shown by the black line~\cite{luk2022a}
}\label{fig:cusp}
\end{figure}

\begin{flushright}
\begin{tabular}{|ll|l|}
\hline
Fig \ref{fig:cusp} \mbox{}& &Moscow University Physics Bulletin. No. 6\,\,- 2024 \\ \hline
to page 6\mbox{} & &To the article by Tribelsky M.I., Luk'yanchuk B.S.\\ \hline
\hline
\end{tabular}\end{flushright}
\newpage
\begin{figure}[tbh!]
\centering
\includegraphics[width=.7\textwidth]{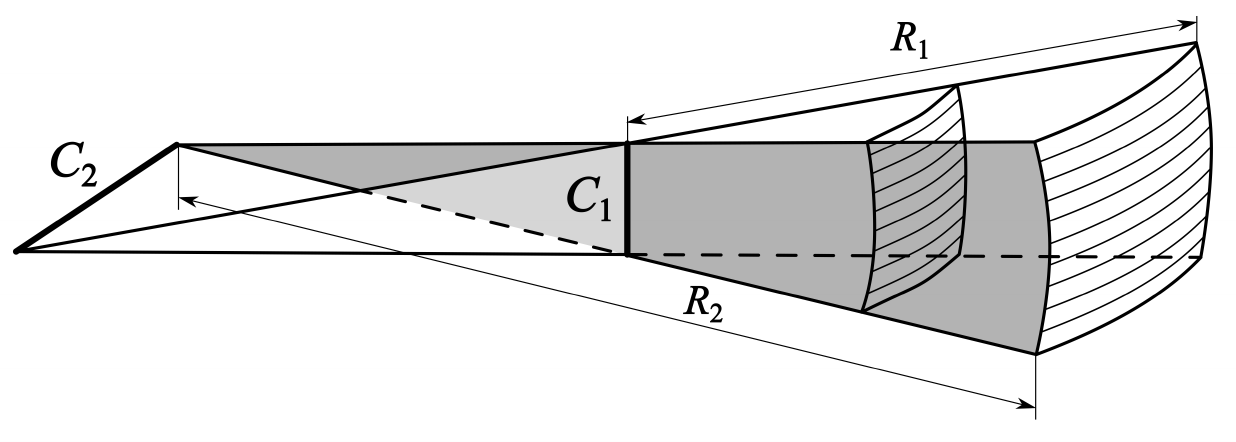}
\caption{Increase in radiation intensity and wavefront curvature as we approach the caustic (schematically)~\cite{Kravtsov1980}. See the text for details}\label{fig:Contraction}
\end{figure}
\begin{flushright}
\begin{tabular}{|ll|l|}
\hline
Fig \ref{fig:Contraction} \mbox{}& &Moscow University Physics Bulletin. No. 6\,\,- 2024 \\ \hline
to page 8\mbox{} & &To the article by Tribelsky M.I., Luk'yanchuk B.S.\\ \hline
\hline
\end{tabular}\end{flushright}
\newpage
\begin{figure}[tbh!]
\centering
\includegraphics[width=\textwidth]{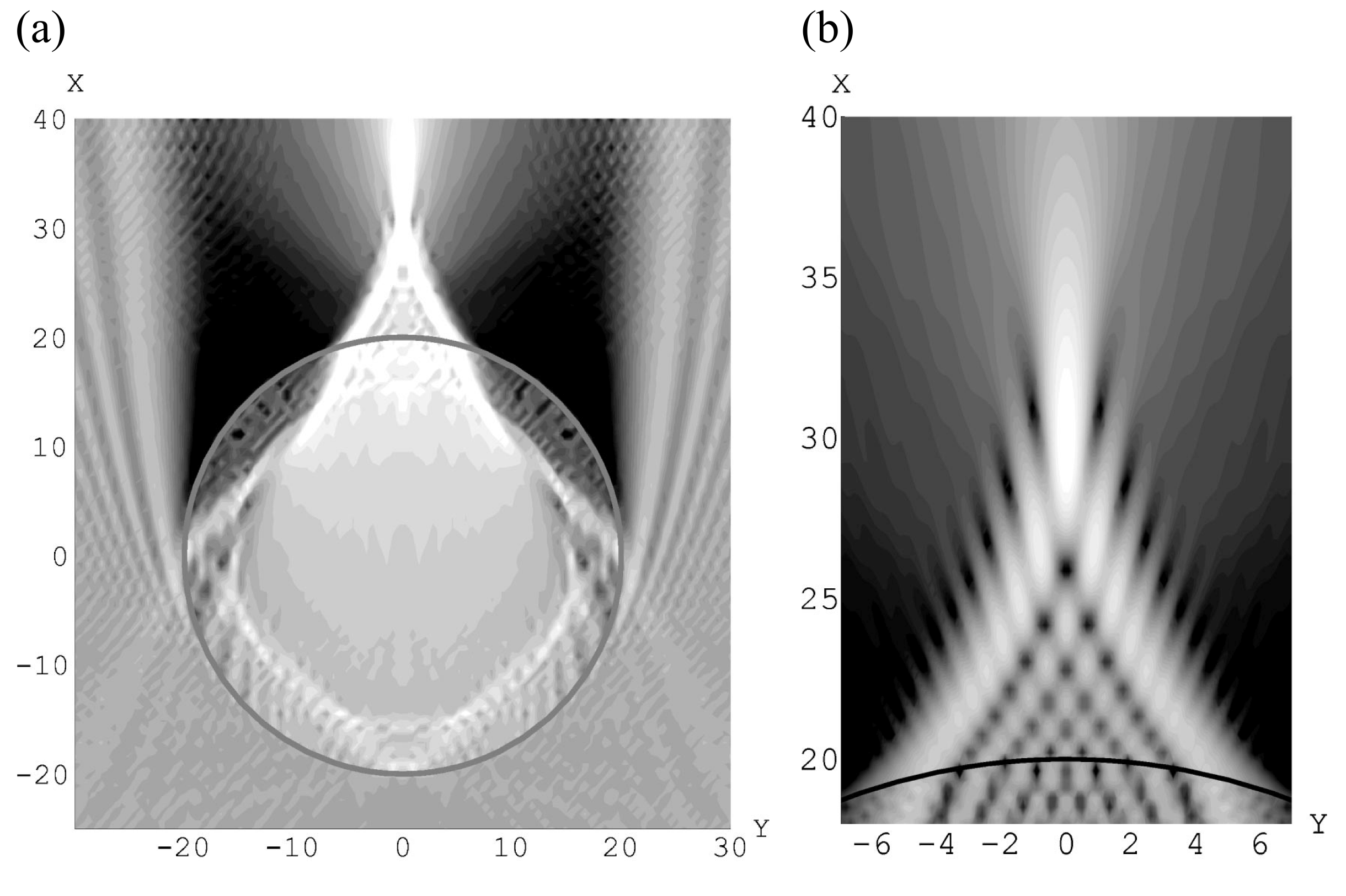}
\caption{Distribution of the intensity $|S|$ of electromagnetic radiation inside the caustic created by scattering of a plane linearly polarized monochromatic wave normally incident on a cylinder with permittivity $\varepsilon=2$ and radius $R = 30\lambda$. Visualization of the exact Mie solution. The field $\vec{E}$ is perpendicular to the plane of the figure (TM polarization), so that the cylinder axis lies in the plane of polarization. (a) --- general diffraction pattern. (b) --- enlarged image of caustics~\cite{zalowich2001a}}\label{fig:Cyl_Cusp}
\end{figure}
\begin{flushright}
\begin{tabular}{|ll|l|}
\hline
Fig \ref{fig:Cyl_Cusp} \mbox{}& &Moscow University Physics Bulletin. No. 6\,\,- 2024 \\ \hline
to page 8\mbox{} & &To the article by Tribelsky M.I., Luk'yanchuk B.S.\\ \hline
\hline
\end{tabular}\end{flushright}
\newpage
\begin{figure}[tbh!]
\centering
\includegraphics[width=.6\textwidth]{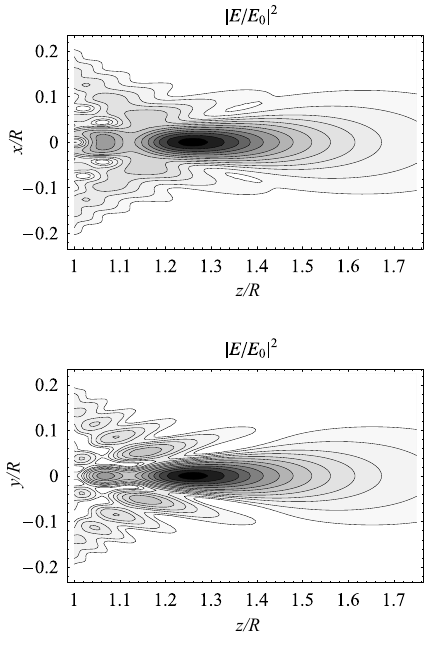}
\caption{Profiles of the field intensity $|E/E_0|^2$ normalized to the incident wave field intensity in the $xz$ and $yz$ planes, when a plane monochromatic linearly polarized electromagnetic wave is focused in a vacuum by a sphere with the refractive index $n=1.5$ and radius $R$. The size parameter $q=kR$ is equal to 100 (in dimensional units, at a radiation wavelength of $\lambda = 248$ nm, such a value of $q$ corresponds to $R \approx 4$ \textmu m). The origin of the coordinate system coincides with the center of the sphere. The incident wave propagates in the positive direction of the $z$-axis, and its vector $\vec{E}$ oscillates along the $x$-axis. In this case, GO gives the position of the geometric focus at the point $z=1.5R$, while the figure shows that, in fact, the position of the intensity maximum is at the point $z \approx 1.25R$~\cite{kofler2006a}} \label{fig:Nikita}
\end{figure}
\begin{flushright}
\begin{tabular}{|ll|l|}
\hline
Fig \ref{fig:Nikita} \mbox{}& &Moscow University Physics Bulletin. No. 6\,\,- 2024 \\ \hline
to page 10\mbox{} & &To the article by Tribelsky M.I., Luk'yanchuk B.S.\\ \hline
\hline
\end{tabular}\end{flushright}
\newpage
\begin{figure}[tbh!]
\centering
\includegraphics[width=.6\textwidth]{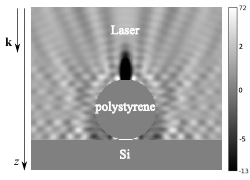}
\caption{Visualization of the exact solution of Maxwell's equations. The structure of the $z$-component of the Poynting vector field $S_z$ in the plane of the equatorial section of a sphere lying on a flat substrate. The field inside the sphere and in the substrate is not shown. A monochromatic plane linearly polarized wave is scattered, the wave vector $\vec{k}$ of which lies in the plane of the figure and is directed vertically downwards. The $z$ axis is parallel to $\vec{k}$, and the plane of polarization coincides with the plane of the figure. $S_z$ is normalized to the intensity of the incident wave. The grayscale is chosen so that black corresponds to the maximum value of the energy flux density directed against the $z\;(S_z<0)$ axis, and white --- to the maximum value of the flux density in the opposite direction $(S_z>0)$. The radiation wavelength is \mbox{$\lambda = 248$ nm}. The material constants correspond to a spherical polystyrene particle {($n=1.6$)} of radius \mbox{0.5 μm} on a silicon substrate ($n=1.57+3.56i$)~\cite{luk2004a}
}\label{fig:OnSubstrate_ Normal}
\end{figure}
\begin{flushright}
\begin{tabular}{|ll|l|}
\hline
Fig \ref{fig:OnSubstrate_ Normal} \mbox{}& &Moscow University Physics Bulletin. No. 6\,\,- 2024 \\ \hline
to page 10\mbox{} & &To the article by Tribelsky M.I., Luk'yanchuk B.S.\\ \hline
\hline
\end{tabular}\end{flushright}
\newpage
\begin{figure}[tbh!]
\centering
\includegraphics[width=\textwidth]{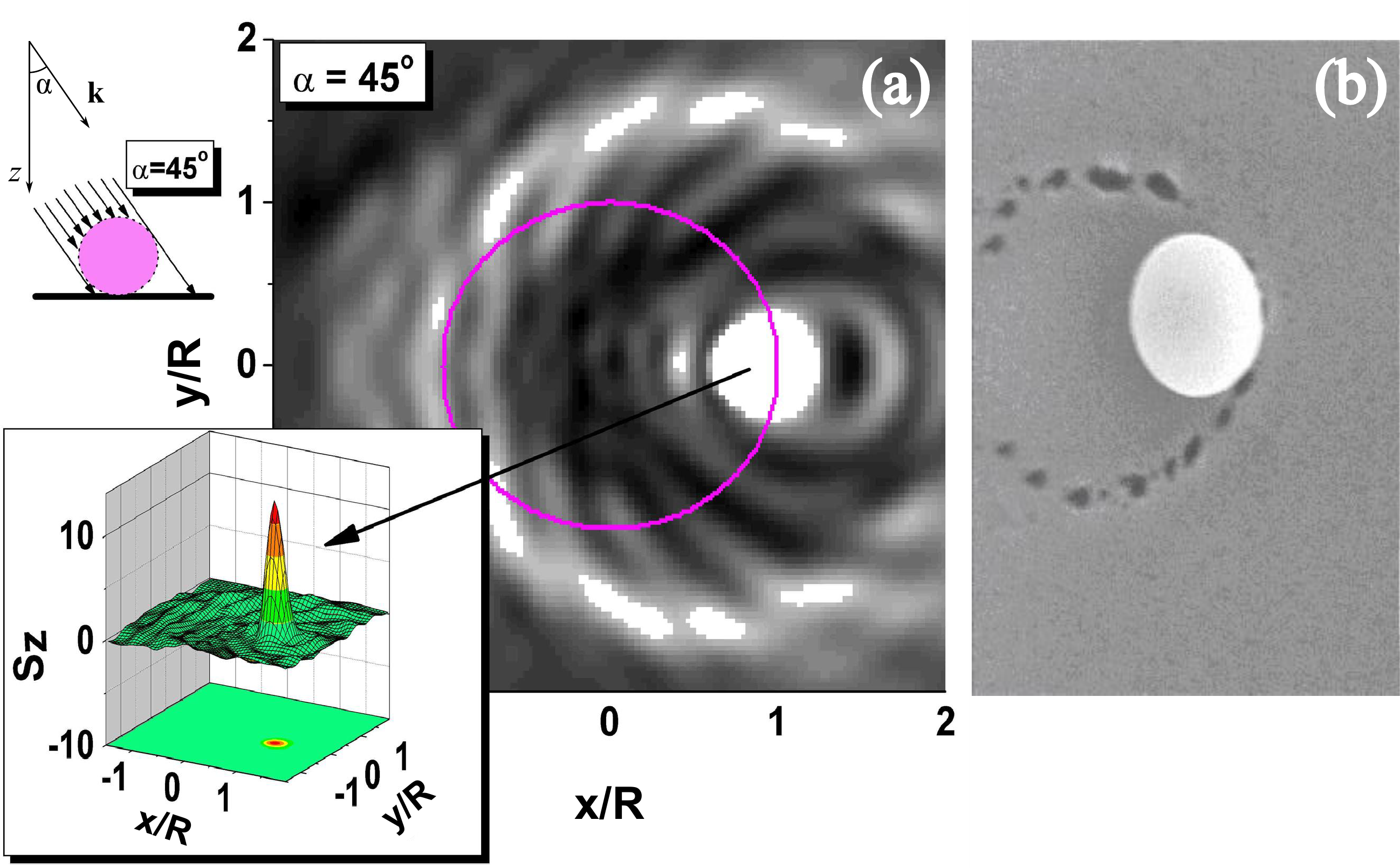}
\caption{(a) --- Calculated distribution of the {$z$-component of the Poynting vector ($S_z$)} on the substrate surface under a polystyrene particle of radius 0.5 \textmu m at an incidence angle of $45^\circ$. White color denotes regions where the absorbed radiation energy exceeds the substrate material damage threshold of 12.0 mJ/cm$^2$. The inset shows the spatial distribution of $S_z$ in the vicinity of the central bright spot located in the geometric shadow region. (b) --- Scanning electron microscopy of the experimentally obtained ``comet pattern'' image on the surface of a $GeSbTe$ film on which a spherical polystyrene particle was located. Irradiation with a pulse of an excimer $KrF$ laser; $\lambda = 248$ nm, pulse duration (FWHM) = 23 ns. The angle of incidence, particle size and optical constants correspond to the calculated data given in panel (a)~\cite{luk2004a}
}\label{fig:Sphere_substrate_45}
\end{figure}
\begin{flushright}
\begin{tabular}{|ll|l|}
\hline
Fig \ref{fig:Sphere_substrate_45} \mbox{}& &Moscow University Physics Bulletin. No. 6\,\,- 2024 \\ \hline
to page {10}\mbox{} & &To the article by Tribelsky M.I., Luk'yanchuk B.S.\\ \hline
\hline
\end{tabular}\end{flushright}
\newpage
\begin{figure}[tbh!]
\centering
\includegraphics[width=.8\textwidth]{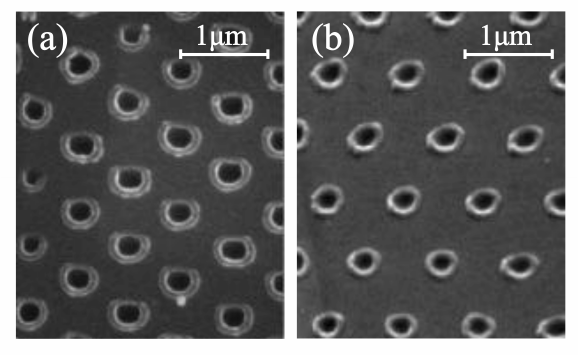}
\caption{Arrays of holes formed by a femtosecond pulsed laser (FWHM = 150 fs, $\lambda = 800$~nm) in a surface coated with a hexagonal monolayer of colloidal spherical $SiO_2$ particles: (a) --- silicon and (b) --- germanium substrates~\cite{munzer2002a}}\label{fig:Holes}
\end{figure}
\begin{flushright}
\begin{tabular}{|ll|l|}
\hline
Fig \ref{fig:Holes} \mbox{}& &Moscow University Physics Bulletin. No. 6\,\,- 2024 \\ \hline
to page {11}\mbox{} & &To the article by Tribelsky M.I., Luk'yanchuk B.S.\\ \hline
\hline
\end{tabular}\end{flushright}
\newpage
\begin{figure}[tbh!]
\centering
\includegraphics[width=.5\textwidth]{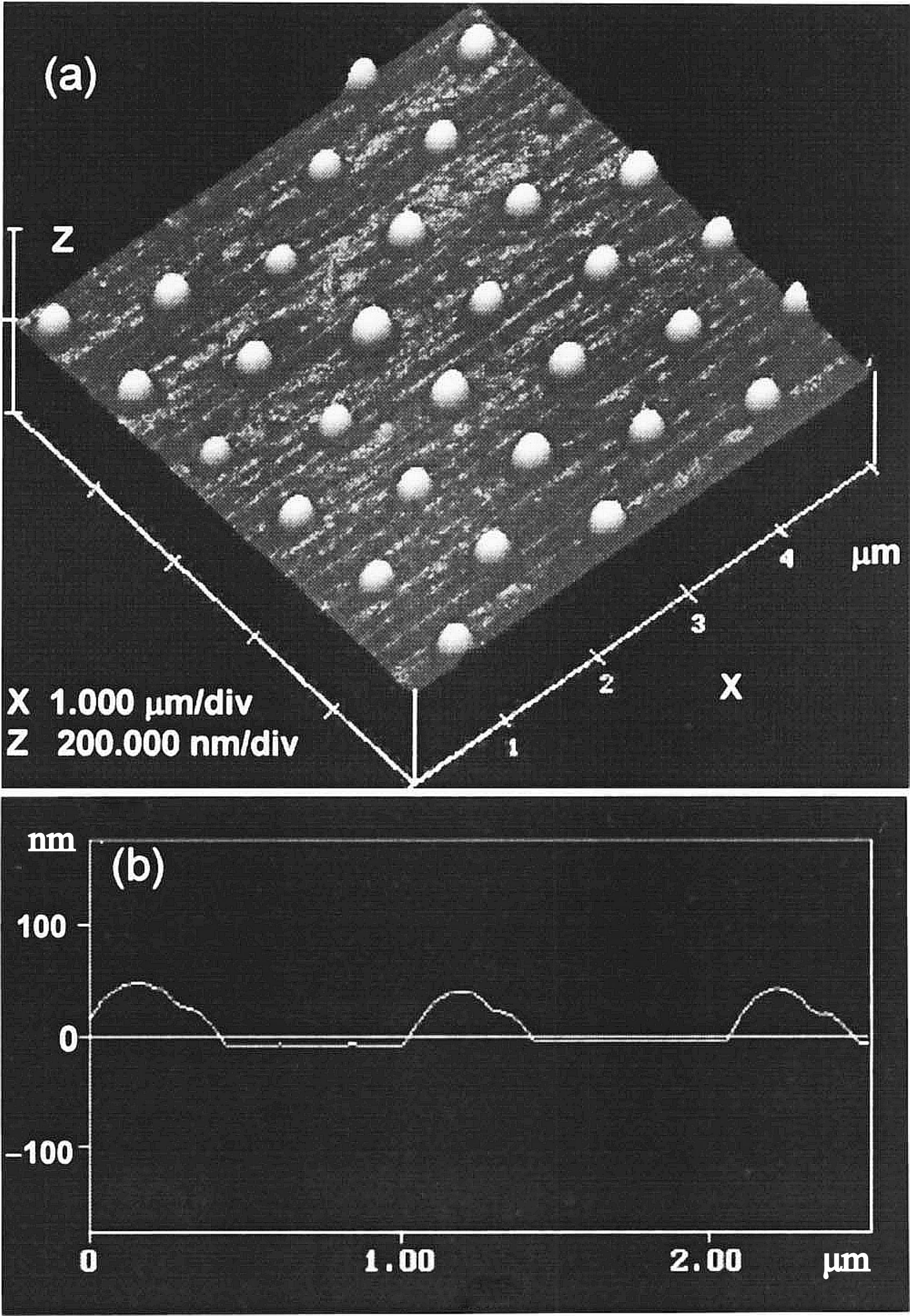}
\caption{(a) --- Silicon surface with nanobumps formed as a result of irradiation of silicon with a {\it KrF\/} laser pulse through an array of dielectric microspheres 1 \textmu m in diameter with a refractive index of $n = 1.6$. (b) --- Profile of this surface in a section passing through the maximum of nanobumps. Zero corresponds to the position of the unirradiated surface. The profiles were obtained using an atomic force microscope~\cite{huang2005nanobump}\label{fig:nanobumps}}
\end{figure}
\begin{flushright}
\begin{tabular}{|ll|l|}
\hline
Fig \ref{fig:nanobumps} \mbox{}& &Moscow University Physics Bulletin. No. 6\,\,- 2024 \\ \hline
to page {11}\mbox{} & &To the article by Tribelsky M.I., Luk'yanchuk B.S.\\ \hline
\hline
\end{tabular}\end{flushright}
\newpage
\begin{figure}[tbh!]
\centering
\includegraphics[width=\textwidth]{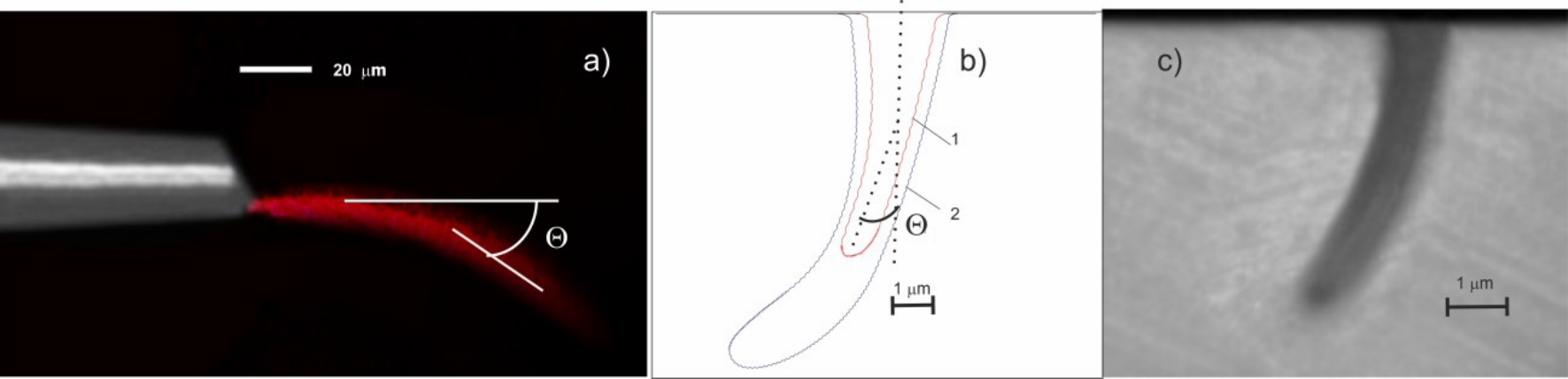}
\caption{{(a) --- Optical fiber with broken symmetry of the end and the formed photonic hook. (b) --- microcrater cross-section profiles for two different laser powers: 1 --- 40 and \mbox{2 --- 80} milliwatts, respectively. (c) --- photograph of a microcrater obtained at a laser power of 40 milliwatts~\cite{hookMinins2024}. Details in the text}}\label{fig:Hook}
\end{figure}
\begin{flushright}
\begin{tabular}{|ll|l|}
\hline
Fig \ref{fig:Hook} \mbox{}& &Moscow University Physics Bulletin. No. 6\,\,- 2024 \\ \hline
to page {12}\mbox{} & &To the article by Tribelsky M.I., Luk'yanchuk B.S.\\ \hline
\hline
\end{tabular}\end{flushright}
\newpage
\begin{figure}[tbh!]
\centering
\includegraphics[width=\textwidth]{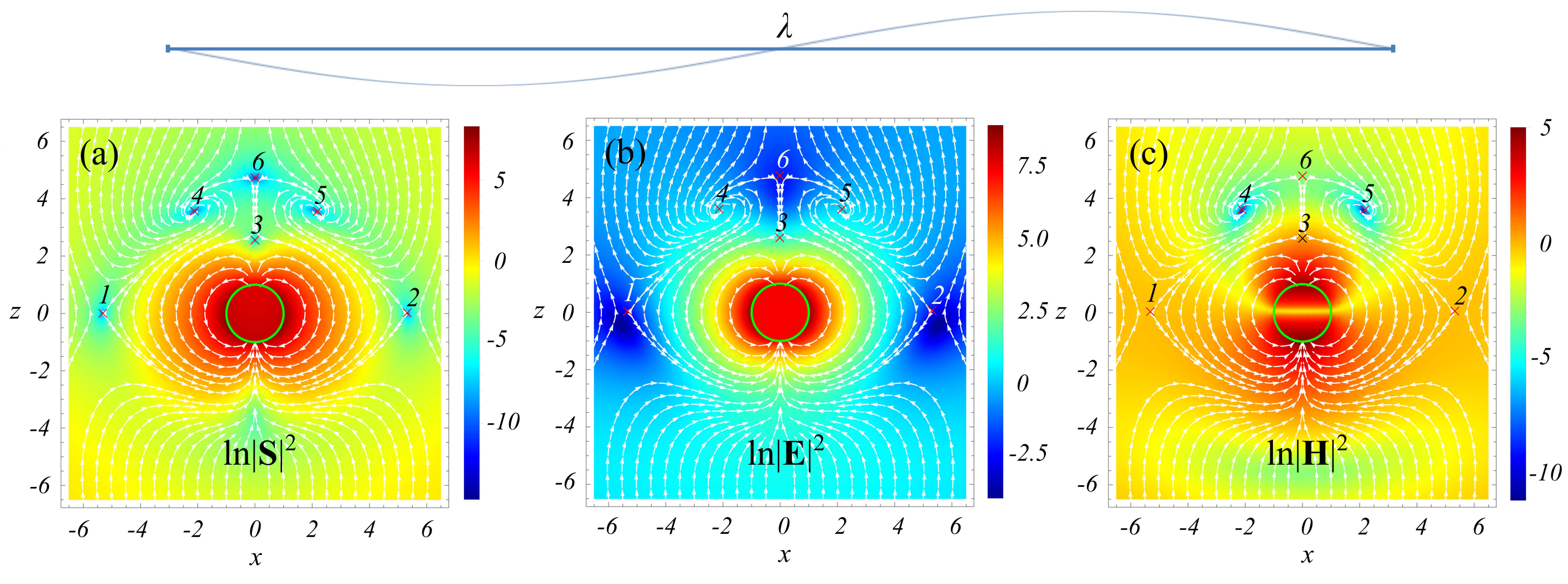}
\caption{The Poynting vector field streamlines, as well as the logarithm of the square of the Poynting vector modulus (a), electric (b) and magnetic (c) fields (shown in color) in the invariant $xz$ plane, calculated using the equation (\ref{eq:S_lines}) based on the exact analytical solution of Maxwell's equations (Mie's solution). The $x$ and $z$ coordinates are normalized to the radius of the sphere $R$. All fields are normalized to the corresponding values in the incident wave. The polarization plane coincides with the plane of the figure. The wave vector $\vec{k}$ of the incident radiation is parallel to the $z$-axis. The size parameter $q=0.3$; $\varepsilon=-2.17$. The surface of the sphere is shown by the solid green line. It follows from the symmetry of the problem that the structures of all fields are symmetric with respect to the plane $x=0$, perpendicular to the plane of the figure. The crosses (x) mark the positions of singular points of the Poynting vector field. Points 4 and 5 are foci, the remaining singularities are saddles; $|\vec{S}|^2=0$ at all singular points. The characteristic scale of the field structure is significantly smaller than the radiation wavelength, the size of which is shown for comparison in the upper part of the figure in the same scale as that for the fields pattern~\cite{Tribelsky2022,TribelskyJETPL}. Details in the text}\label{fig:Sphere_lines}
\end{figure}
\begin{flushright}
\begin{tabular}{|ll|l|}
\hline
Fig \ref{fig:Sphere_lines}\mbox{}& &Moscow University Physics Bulletin. No. 6\,\,- 2024 \\ \hline
to page {21}\mbox{} & &To the article by Tribelsky M.I., Luk'yanchuk B.S.\\ \hline
\hline
\end{tabular}\end{flushright}
\newpage
\begin{figure}[tbh!]
\centering
\includegraphics[width=0.3\textwidth]{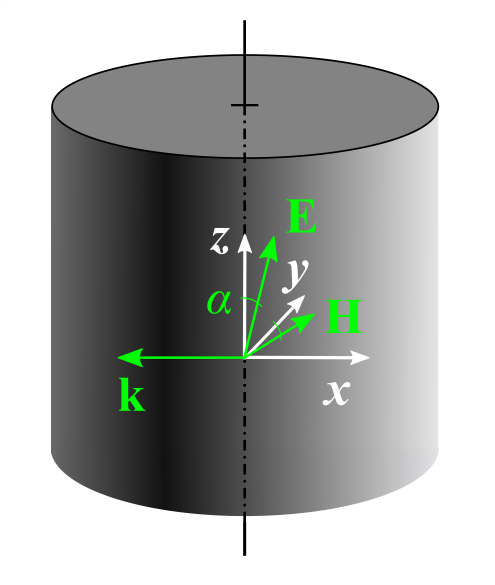}
\caption{Relative orientation of the cylinder, coordinate system and vectors $\vec{k}$, $\vec{E}$ and $\vec{H}$ of the incident wave; $\alpha$ is the angle between axis $z$ and vector $\vec{E}$, equal to angle between axis $y$ and vector $\vec{H}$}
\label{fig:Oblique}
\end{figure}
\begin{flushright}
\begin{tabular}{|ll|l|}
\hline
Fig \ref{fig:Oblique} \mbox{}& &Moscow University Physics Bulletin. No. 6\,\,- 2024 \\ \hline
to page {25}\mbox{} & &To the article by Tribelsky M.I., Luk'yanchuk B.S.\\ \hline
\hline
\end{tabular}\end{flushright}
\newpage
\begin{figure}[tbh!]
\centering
\includegraphics[width=\textwidth]{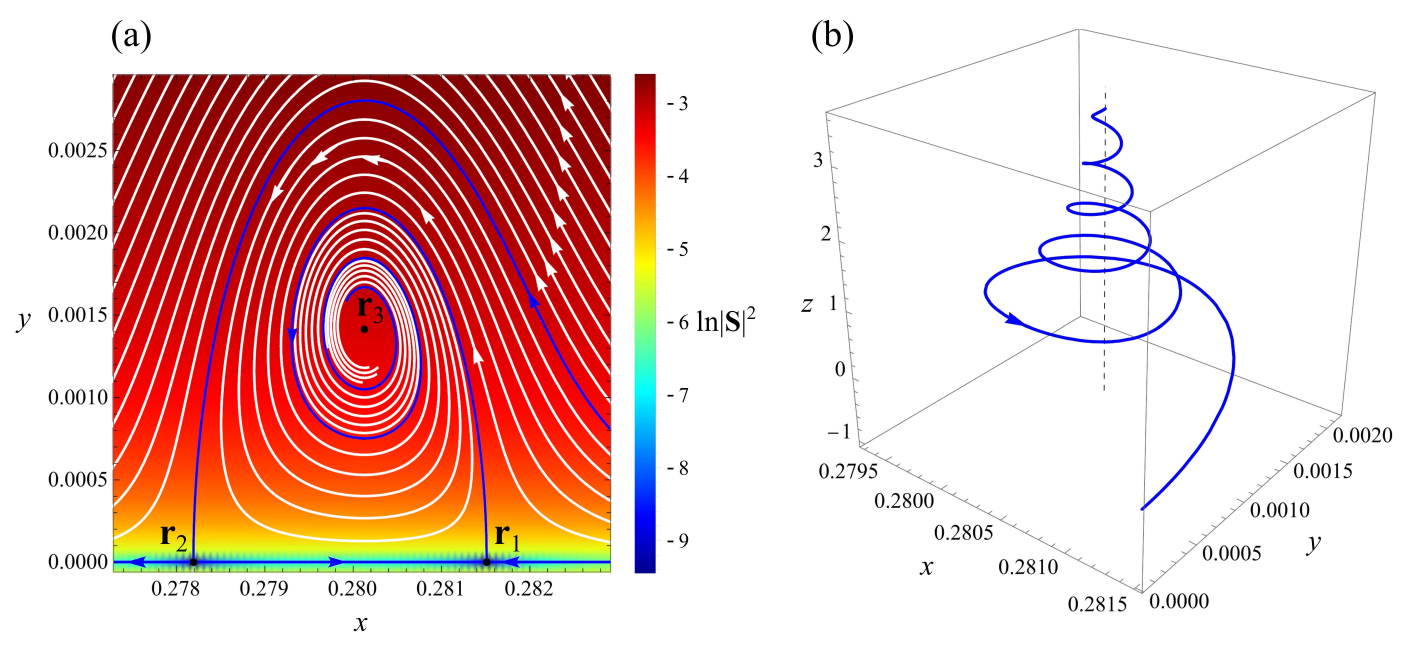}
\caption{The Poynting vector field inside an infinite circular cylinder, normalized to the incident wave intensity, calculated based on the exact solution of Maxwell's equations; \mbox{$\varepsilon = 17.775+0.024 i$}, which corresponds to the permittivity of germanium at a wavelength of 1590 nm~\cite{Polyanskiy}; $\alpha = 45.403^\circ$. {Spatial coordinates are normalized to the cylinder radius $R$. } The two-dimensional projection (a) of the field onto the plane perpendicular to the cylinder axis contains three singularities, indicated by black dots: two saddles ($r_{1,2}$) lying on the $x$-axis and a stable focus (${r}_3$) that does not belong to this axis. At the same time, in three-dimensional space, the saddles $r_{1,2}$ are true singularities, while the point $r_3$ becomes regular. This is clearly seen in panel (b), where in three-dimensional space a segment of the separatrix whisker is shown, emerging from the saddle ${r}_1$ and winding around a vertical straight line (shown by the dotted line), whose projection onto the $xy$ plane corresponds to the {point} ${r}_3$~\cite{tribelsky2024_LPR,Tribelsky2023}. Note the difference in the scale of the axes in \mbox{panel (b)}}\label{fig:2d-3D}
\end{figure}

\begin{flushright}
\begin{tabular}{|ll|l|}
\hline
Fig {\ref{fig:2d-3D}} \mbox{}& &Moscow University Physics Bulletin. No. 6\,\,- 2024 \\ \hline
to page {26}\mbox{} & &To the article by Tribelsky M.I., Luk'yanchuk B.S.\\ \hline
\hline
\end{tabular}\end{flushright}

%\newpage
\end{document}